\documentclass[11pt]{article}
\usepackage{caption}
\usepackage{scalerel}
\usepackage{amsthm,latexsym,amsxtra,graphicx,appendix,epstopdf,feynmf,hyperref,setspace,fix-cm}
\usepackage[normalem]{ulem}
\usepackage{makeidx}
\usepackage{fullpage}
\usepackage{amsmath}
\usepackage{amssymb}
\usepackage{setspace}
\usepackage{bbm}
\usepackage{dsfont}
\usepackage{graphics}
\usepackage{epsfig}
\usepackage[font=footnotesize,labelfont=bf,justification=centerlast,width=.94\textwidth]{caption}
\usepackage{cite}
 \usepackage{multirow}
\usepackage{array,booktabs}
\usepackage{subfigure}
\usepackage{color}
\usepackage[makeroom]{cancel}
\usepackage{tikz}
\usepackage{pgfplots}
\usetikzlibrary{intersections, pgfplots.fillbetween}
\usepackage{tensor}
\usepackage{simplewick}
\usepackage{frcursive}
\usepackage{soul}

\usepackage{caption}
\usepackage{hyperref}

\hypersetup{
    bookmarks=true,%
    colorlinks,%
    citecolor=blue,%
    filecolor=blue,%
    linkcolor=blue,%
    urlcolor=blue
}
\usepackage{float}
\usepackage{slashed}
\usepackage{tikz}
\usetikzlibrary{decorations.pathmorphing}

\newcommand{\bi}{\begin{itemize}}
\newcommand{\ei}{\end{itemize}}
\newcommand{\bea}{\begin{eqnarray}}
\newcommand{\eea}{\end{eqnarray}}
\newcommand{\be}{\begin{equation}}
\newcommand{\ee}{\end{equation}}

\def\XXint#1#2#3{{\setbox0=\hbox{$#1{#2#3}{\int}$}
     \vcenter{\hbox{$#2#3$}}\kern-.5\wd0}}

\numberwithin{equation}{section}
\allowdisplaybreaks  

\begin{document}

\vspace*{2.5cm}
\begin{center}
{ \Large \textsc{Renormalisation Group Flows of  Deformed SYK Models}} \\ \vspace*{1.3cm}

\end{center}

\begin{center}
Dionysios Anninos, Dami\'an A. Galante, and Sameer U. Sheorey  \\ 
\end{center}
\begin{center}
{
\footnotesize
Department of Mathematics, King's College London, Strand, London WC2R 2LS, UK
}
\end{center}
\begin{center}
{\textsf{\footnotesize{
dionysios.anninos@kcl.ac.uk, damian.galante@kcl.ac.uk, sameer.sheorey@kcl.ac.uk}} } 
\end{center}

\vspace*{0.5cm}

\vspace*{1.5cm}
\begin{abstract}
\noindent
\\ \\ 

We explore computationally tractable deformations of the SYK model. The deformed theories are described by the sum of two SYK Hamiltonians with differing numbers, $q$ and $\tilde{q}$, of interacting fermions. In the large $N$ limit, employing analytic and numerical tools, we compute finite temperature correlation functions and thermodynamic quantities. We identify a novel analytically solvable model in the large $q$ limit. We find that, under certain circumstances, the thermal RG flow in the strongly coupled infrared phase exhibits two regions of linear-in-temperature entropy, which we interpret in terms of Schwarzian actions. Using conformal perturbation theory we compute the leading relevant correction away from the intermediate near-conformal fixed point. Holographic spacetimes in two spacetime dimensions that reproduce the thermodynamics of the microphysical theory are discussed. These are flow geometries that interpolate between two Euclidean near-AdS$_2$ spacetimes with different radii. The Schwarzian soft mode corresponding to the AdS$_2$ region in the deep interior resides entirely within the geometric regime.

\end{abstract}




\newpage

\tableofcontents

\section{Introduction}
\label{section: Introduction}

Given the description of a theory at a conformally invariant fixed point, one is naturally led to examine deformations causing the theory to flow toward a novel phase in the infrared. Sufficiently close to the fixed point, one can quantify the deformations by the set of primary operators which are relevant with respect to the original fixed point. The richer the space of relevant operators, the more elaborate the landscape of renormalisation group (RG) flows away from the underlying fixed point, and the more ample the opportunity to design particular infrared behaviour.

A useful strategy to gain insight into RG flows is to study the theory at finite temperature and use this as the energy scale of the problem \cite{CASTRONETO1993525, Zabzine:1997gh, Appelquist:1999hr,Anninos:2020cwo,Delacretaz:2021ufg}. In this work we will take this approach to analyse the effect of a relevant deformation causing a flow away from the near-fixed point of the Sachdev-Ye-Kitaev (SYK) model \cite{Sachdev:1992fk,Kitaev:2017awl,Maldacena:2016hyu}. The structure of the flow will be revealed through the detailed dependence of thermal correlations and thermodymanic quantities on the strength of the relevant term. The presence of additional fixed points and other properties of the flow are revealed through such physical quantities.

The SYK model is a theory of $N$ interacting Majorana fermions subject to randomly disordered couplings. The type of deformation we consider is itself disordered, and further to this, we study the problem at both vanishing and finite temperature. The essential motivation behind our work is to develop a new direction in the study of holographic renormalisation \cite{deBoer:1999tgo,deHaro:2000vlm} by identifying tractable renormalisation group flows for strongly coupled theories at large $N$. From the perspective of the gravitational description, the renormalisation group flow manifests itself in a geometry that flows away from the asymptotically AdS boundary describing the fixed point. The microphysical flow imposes, directly from its underlying quantum consistency conditions, constraints on the space of deformed holographic bulk theories whose description is often restricted to low energy effective field theory. 

The basic challenge is that strongly coupled fixed points with tractable renormalisation group flows are hard to come across. To address this challenge, we consider the SYK model whose strongly coupled low temperature phase has been argued to exhibit holographic properties \cite{Maldacena:2016upp,Jensen:2016pah, Engelsoy:2016xyb} at large $N$. Although relevant deformations of SYK have not been studied extensively in the literature, there are exceptions \cite{Garcia-Garcia:2017bkg,Jiang:2019pam,Lunkin:2020tbq,Anninos:2020cwo,Nandy:2022hcm}. Moreover, there have been a host of interesting variations  of SYK including entangling a pair of SYK theories to each other \cite{Maldacena:2018lmt,Garcia-Garcia:2019poj}, endowing  SYK type models with internal global symmetries \cite{Gross:2016kjj,AnninosMarginal,Yoon:2017nig,Gu:2019jub}, non-Hermitian SYK Hamiltonians modelling open quantum systems \cite{Liu:2020fbd,Garcia-Garcia:2021rle}, models of SYK chains and higher-dimensional analogues \cite{Gu:2016oyy,Goel:2018ubv}, and supersymmetric extensions \cite{Anninos:2016szt,Fu:2016vas}. In this work we employ a variety of analytical and numerical techniques 
to analyse a class of tractable strongly coupled renormalisation group flows away from the near-fixed point of SYK for a variety of deformations. Concretely, we examine the Hamiltonian
\begin{equation}
    H_{q} = i^{\frac{q}{2}} \sum_{1\leq i_1 < i_2 <\cdot\cdot\cdot< i_q\leq N} J_{i_1i_2\cdot\cdot\cdot i_q}\psi_{i_1}\psi_{i_2}\cdot\cdot\cdot \psi_{i_q}~, \quad\quad q \in 2 \mathbb{Z}^+ ~,
\end{equation}
deformed by the operator $s\, H_{\tilde{q}}$ with $\tilde{q}<q$, where $s$ is a dimensionless coupling and $\psi_i$ are $N$ Majorana fermions. The couplings of both $H_q$ and $H_{\tilde{q}}$ are drawn from a Gaussian ensemble. The deformation is implemented at the level of the ultraviolet degrees of freedom. Nonetheless, concrete evidence is provided that for sufficiently small $s$, the deformation can be viewed as a relevant deformation by a specific conformal operator of the near-fixed point describing the low energy physics of the undeformed SYK model. Previous work \cite{Anninos:2020cwo} has established this in the large $q$ limit with $q/\tilde{q} = 2$. Here, we establish this phenomenon at both large and finite $q$, $\tilde{q}$. Moreover, the effect is seen for several values for $ n \equiv q/\tilde{q}$. The flow is shown to end at a near-fixed point in the deep infrared, where the theory is captured by an SYK theory governed by $H_{\tilde{q}}$. Interestingly, the Schwarzian sector of the theory in the deep infrared resides entirely within the strongly coupled sector of the theory. From a holographic point of view, this can be viewed as a soft mode emerging in the interior of a bulk asymptotically AdS$_2$ spacetime. 

The paper is structured as follows. In section \ref{sec: Brief review of the SYK model} we briefly review the SYK model for general $q$. We discuss the large $N$ saddle-point Schwinger-Dyson equations and the large $q$ limit. In section \ref{The deformed SYK model} we introduce the deformations of interest, and the corresponding large $N$ saddle-point Schwinger-Dyson equations. The theory is considered at finite temperature. In section \ref{sec_deep_IR} we study the low temperature behaviour of the deformed models, as well as providing evidence for the existence of an intermediate near-conformal fixed point for a subclass of these models. When possible, we consider analytical expressions, including  a new regime of small $\varepsilon \equiv n-1$ permitting analytic treatment. In section \ref{sec_int_IR} we explore the structure of the renormalisation group flow in the vicinity of each near-fixed point and uncover the soft-mode theories governing the leading thermodynamic behaviour. We also show that conformal perturbation theory can be applied to study the leading relevant deformation away from the intermediate near-fixed point. In the outlook section \ref{sec_outlook} we discuss how our results can be interpreted from a holographic point of view, in the form of a JT gravity theory with deformed dilaton potential. 


\section{Brief review of the SYK model}
\label{sec: Brief review of the SYK model}
The SYK model is a quantum mechanical model  with random all-to-all interactions. The observables of the theory are built from $N$ Majorana fermions, $\psi_{i}$, that obey equal time anti-commutation relations
\begin{equation}
\label{Majorana anticom relns}
    \{\psi_i,\psi_j\} = \delta_{ij}\;, \quad\quad i,j=1,\ldots,N~.
\end{equation}
The Hamiltonian of the model is given by
\begin{equation}
\label{SYK Hamiltonian}
    H_{q} = (i)^{\frac{q}{2}} \sum_{1\leq i_1 < i_2 <\ldots< i_q\leq N} J_{i_1i_2\ldots i_q}\psi_{i_1}\psi_{i_2}\ldots \psi_{i_q}~, \quad\quad q \in 2 \mathbb{Z}^+ ~,
\end{equation}
where the coupling constants of the theory are all independently drawn from the same probability distribution that satisfies
\begin{equation}
\label{coupling distribution}
    \langle J_{i_1i_2\cdot\cdot\cdot i_q}\rangle = 0~, \quad\quad \langle J^{2}_{i_1i_2\cdot\cdot\cdot i_q}\rangle = \frac{2^{q-1}}{q}\frac{\mathcal{J}^2 (q-1)!}{N^{q-1}}~.
\end{equation}
The dimensionality of the Hilbert space is $2^{N/2}$ and the theory is numerically amenable to exact diagonalisation procedures for reasonably large values of $N$.\footnote{In \cite{Cotler:2016fpe}, for instance, results for $N=34$ are reported. For certain observables, it is also possible to partially diagonalise the Hamiltonian using Krylov methods to get up to $N=60$, see \cite{Kobrin:2020xms}.} A review of the SYK model can be found in \cite{Sarosi:2017ykf,Rosenhaus:2018dtp}, among other articles.

\subsection{Large $N$ limit}

From the perspective of the path integral, it is useful to express the theory in terms of bi-local fields $G(\tau_1,\tau_2)$, $\Sigma(\tau_1,\tau_2)$ \cite{Sachdev:2015efa, Maldacena:2016hyu,Kitaev:2017awl}. The Euclidean time coordinate $\tau \sim \tau + \beta$ is periodically identified with period given by the inverse temperature $\beta$. Physically, $G(\tau_1,\tau_2)$ computes the (time-ordered) thermal two point function 
\begin{equation}
G(\tau_1,\tau_2) = \frac{1}{N}  \sum_{i=1}^{N} \langle T\psi_i(\tau_1)\psi_i(\tau_2) \rangle~.
\end{equation}
In terms of $G$ and $\Sigma$ the  action reads
\begin{equation} \label{G Sigma action}
I = -\frac{1}{2}\log\det\left(\delta(\tau_1-\tau_2)\partial_{\tau_2}-\Sigma(\tau_1,\tau_2)\right) + \frac{1}{2}\int_{0}^{\beta}\int_{0}^{\beta}d\tau_1 d\tau_2\left(\Sigma(\tau_1,\tau_2) G(\tau_1,\tau_2) - \mathcal{J}^2\frac{2^{q-1}}{q^2}G(\tau_1,\tau_2)^q\right)~,
\end{equation}
and the disorder averaged partition function of the theory is given by
\begin{equation}
\label{effective partition function PI}
\langle Z(\beta)\rangle_J = \int [DGD\Sigma]e^{-NI[G,\Sigma]}~,
\end{equation}
where we indicate a disorder average by $\langle \bullet \rangle_J$. At large $N$, the theory permits a saddle point approximation. The resulting Schwinger-Dyson equations are  the following integro-differential equations
    \begin{eqnarray}\label{field equations large N}
        G^{-1}(\tau_1,\tau_2) &=&  \delta(\tau_1-\tau_2)\partial_{\tau_2}-\Sigma(\tau_1,\tau_2) ~, \\
       \Sigma(\tau_1,\tau_2) &=& \frac{2^{q-1}}{q} \mathcal{J}^2 G(\tau_1,\tau_2)^{q-1}~. \label{field equations large N ii}
    \end{eqnarray}
The above equations can be solved numerically using a recursive algorithm and the fast Fourier transform \cite{Maldacena:2016hyu}. In the IR of the theory, given by $|\tau_1 -\tau_2| \gg 1/\mathcal{J}$, we can self-consistently drop the $\delta(\tau_1-\tau_2)\partial_{\tau_2}$ term in \eqref{field equations large N}, resulting in an effective theory described by the equations
\begin{eqnarray}\label{IR field equations}
\int_{0}^{\beta} d\tau'\; G(\tau_1,\tau')\Sigma(\tau',\tau_2) &=& -\delta(\tau_1-\tau_2)~,\\
\Sigma(\tau_1,\tau_2) &=& \frac{2^{q-1}}{q} \mathcal{J}^2G(\tau_1,\tau_2)^{q-1}~.\label{IR field equations ii}
\end{eqnarray}
Provided $\Delta=1/q$ these above equations are invariant under the transformations
\begin{eqnarray}
\label{reparametrisation symmetry}
    G(\tau_1,\tau_2) \to \tilde{G}(\tau_1,\tau_2) &=&  \phi'(\tau_1)^{\Delta} \, G(\phi(\tau_1),\phi(\tau_2)) \, \phi'(\tau_2)^{\Delta}~, \\
    \Sigma(\tau_1,\tau_2) \to \tilde{\Sigma}(\tau_1,\tau_2) &=&  \phi'(\tau_1)^{\Delta(q-1)} \, \Sigma(\phi(\tau_1),\phi(\tau_2))  \, \phi'(\tau_2)^{\Delta(q-1)}~,
\end{eqnarray}
with $\phi(\tau)$ a smooth, monotonically increasing function that maps the thermal circle to the thermal circle with single unit of winding. The structure of $\phi(\tau)$ is that of a reparameterisation of the circle to itself.

In the IR, the SYK model is approximated by a one-dimensional conformal field theory \cite{Kitaev:2017awl, Maldacena:2016hyu}. The fermions $\psi_i$ transform as primary operators of conformal weight $\Delta = 1/q$. At the level of the action, the low-energy effective description is given by 
\begin{equation}
\label{CFT action}
    I_{\text{CFT}} = -\frac{1}{2}\log\det(-\Sigma(\tau_1,\tau_2) ) + \frac{1}{2}\int_0^\beta\int_0^\beta d\tau_1 d\tau_2\left(\Sigma(\tau_1,\tau_2) G(\tau_1,\tau_2) - \mathcal{J}^2\frac{2^{q-1}}{q^2}G(\tau_1,\tau_2)^q\right)~.
\end{equation}
 The solution to the IR Schwinger-Dyson equations \eqref{IR field equations} and \eqref{IR field equations ii} is given by
\begin{equation}
\label{full two point on circle}
    G_{\phi}(\tau_1,\tau_2) =  \phi'(\tau_1)^{\Delta} \, b \, \mbox{sgn}(\tau_1-\tau_2)  \left(\frac{\pi}{\beta \mathcal{J} \sin\left(\frac{\pi(\phi(\tau_1)-\phi(\tau_2))}{\beta}\right)}\right)^{2\Delta}\,\phi'(\tau_2)^{\Delta}~,
\end{equation}
with 
\begin{equation}\label{constant b}
    b = \frac{1}{2} \left(\frac{(1-2 \Delta ) \tan (\pi  \Delta )}{\pi  \Delta }\right)^{\Delta }~.
\end{equation}
All solutions $G_{\phi}$ have the same action when evaluated on the conformal action (\ref{CFT action}). As such, the saddle approximation naively diverges as the volume of the reparameterisation group. To get a finite answer we must account for the effect of the leading `irrelevant' correction away from the conformal action. It is given by the Schwarzian action \cite{Maldacena:2016hyu}
\begin{equation}
\label{Scwharzian action}
    I_{\text{Sch}} = -\frac{\alpha(q)}{2\mathcal{J}}\int_{0}^{\beta} d\tau\;\left(\left(\frac{2\pi}{\beta}\right)^{2}\phi'(\tau)^{2} - \left(\frac{\phi''(\tau)}{\phi'(\tau)}\right)^{2}\right)~.
\end{equation}
The constant $\alpha(q)$ has to be determined numerically by solving the full Schwinger-Dyson equations, as discussed further in Appendix \ref{app: alpha}, as its precise value does not follow from IR considerations. The Schwarzian action explicitly breaks the reparametrisation symmetry of the conformal action down to an unphysical $SL(2,\mathbb{R})$ reparametrisation group. The final path integral must still be divided by the volume of the residual $SL(2,\mathbb{R})$ to be made sense of \cite{Stanford:2017thb, Anninos:2021ydw}.

Given the Schwarzian theory \eqref{Scwharzian action}, one can compute thermodynamic quantities to leading order in the saddle point approximation. For instance, given the on-shell solution $\phi(\tau) = \tau$, the free energy $F_{\text{Sch}}$ is found by taking the Schwarzian action on shell
\begin{equation}
\label{finite q specific heat}
 -   \frac{\beta F_{\text{Sch}}}{N} = \frac{2\pi^{2} \alpha(q)}{\beta \mathcal{J}}~,
\end{equation}
and is found to be quadratic in the temperature. Given an expression for free energy $F$, the thermodynamic entropy $S$ can be computed  as
\begin{equation}
\label{S from F}
    S = (1-\beta\partial_{\beta})(-\beta F)~.
\end{equation}
It is straightforward from (\ref{finite q specific heat}) to verify that the entropy of the Schwarzian theory is linear in the temperature,
\begin{equation} \label{entropy sch}
\frac{S_{\text{Sch}}}{N} = \frac{4 \pi^2 \alpha(q)}{\beta \mathcal{J}} \,.
\end{equation}
Additionally, the zero temperature entropy of the SYK can be computed explicitly \cite{Kitaev:2017awl, Maldacena:2016hyu} such that the entropy of the SYK model admits the following small temperature expansion
\begin{equation}
\label{finite q entropy}
    \frac{S}{N} = \left(S^{\mathrm{free}}_0 - \int_{0}^{1/q} d x\; \pi\left(\frac{1}{2}-x\right) \tan \pi x \right) +  \frac{4\pi^{2}\alpha(q)}{\beta \mathcal{J}} + \cdots~,
\end{equation}
where $S^{\mathrm{free}}_0 \equiv  {\log 2}/{2}$ is the zero temperature entropy of a free fermion. 

\subsection{Large $q$ limit}
The SYK model admits further computational control if, after taking the large $N$ limit, we also take the large $q$ limit.\footnote{Another solvable case is known as the double-scaled SYK model, obtained by taking both the large $N$ and large $q$ limit, but with $N/q^2$ fixed. See, for instance, \cite{Berkooz:2018qkz, Berkooz:2018jqr}.} In this case, we can expand the two-point function $G(\tau_1,\tau_2) = G(\tau_1-\tau_2)$ as
\begin{equation}
\label{large q two point}
    G(\tau) = \frac{\mbox{sgn}(\tau)}{2}\left(1 + \frac{g(\tau)}{q}+\mathcal{O}(1/q^{2})\right)~.
\end{equation}
To leading order in $q$, the Schwinger-Dyson equations (\ref{field equations large N}) and (\ref{field equations large N ii})  become a single ordinary differential equation for $g(\tau)$, namely
\begin{equation}
    \partial_{\tau}^{2}g(\tau) =  2\mathcal{J}^2e^{g(\tau)}~. 
    \end{equation}
Supplemented by thermal boundary conditions, $g(0) = g(\beta) = 0$, this equation can be solved analytically and yields,
\begin{equation}
e^{g(\tau)} = \frac{\cos^2 \nu}{\cos^2 \left( 2 \nu \left( \frac{1}{2} - \frac{|\tau|}{\beta} \right)\right)}\, , \quad\quad \beta \mathcal{J} = \frac{2 \nu}{\cos \nu} \,.
\end{equation}
Given $g(\tau)$, we can compute the complete thermodynamics of the theory by evaluating the action (\ref{G Sigma action}) on-shell to leading order in the large $q$ expansion,
\begin{equation}
\label{large q free energy single SYK}
\frac{\beta F}{N} = -S^{\mathrm{free}}_0 - \frac{\beta}{8 q^{2}} \int_{0}^{\beta} d \tau\left[\frac{1}{2}\left(\partial_{\tau} g(\tau)\right)^{2}+ 2 \mathcal{J}^{2} e^{g(\tau)} \right] + \cdots ~.
\end{equation}
For large $\beta \mathcal{J}$, we obtain that the entropy at large $q$ is given by
\begin{equation}
\label{entropy large q}
    \frac{S}{N} = \left(S^{\mathrm{free}}_0 -\frac{\pi^2}{4 q^2}\right) + \frac{\pi^2}{q^2}\frac{1}{\beta \mathcal{J}} + \cdot\cdot\cdot ~.
\end{equation}
By comparing (\ref{entropy large q}) with (\ref{entropy sch}), we see that $\alpha(q) \to 1/4q^2$ as $q \to \infty$. Next order corrections in the large $q$ limit have been studied in \cite{Tarnopolsky:2018env}.

\section{Deformed SYK models}
\label{The deformed SYK model}
In this section we  introduce a family of deformations of the single SYK in which the Hamiltonian is the sum of two SYK Hamiltonians with different numbers of fermions in the interactions. The behaviour of the deformed models can be thought of in terms of an RG flow of the original SYK theory. One way of studying the model at different energy scales is to turn on the temperature. The deformed models can be solved either numerically or analytically for a wide range of parameters of the theory. At finite $q$, aspects of these models have been studied in \cite{Garcia-Garcia:2017bkg, Lunkin:2020tbq, Nandy:2022hcm}, while at large $q$ analytically tractable examples have been considered in \cite{Jiang:2019pam, Anninos:2020cwo}. 

\subsection{Deformed Hamiltonian and effective action}

The Hamiltonian of the deformed SYK models is given by
\begin{equation}
\label{deformed Hamiltonian}
    H_{\mathrm{def}}=H_{q}+s H_{\tilde{q}}  ~,
\end{equation}
where $s$ is a  tuneable dimensionless parameter and the Hamiltonian $H_x$ denotes the Hamiltonian (\ref{SYK Hamiltonian}) of a single SYK model with $x$ fermion interactions. We will assume that $q\geq\tilde{q}$. Unitarity imposes that $s \in 
\mathbb{R}$, and without loss of generality we can further restrict to $s \in \mathbb{R}^+$. The term $s H_{\tilde{q}}$ can be viewed as a relevant deformation of the model $H_{q}$ that induces an RG flow and modifies the thermodynamic behaviour of the model in the infrared. 

Similar to the single SYK case, in the large $N$ limit, the deformed action can be described in terms of bi-local fields  \cite{Jiang:2019pam, Anninos:2020cwo} 
\begin{equation}
\label{deformed bilocal action}
I = -\frac{1}{2}\log\det(\partial_{\tau}-\Sigma) + \frac{1}{2}\int_0^\beta\int_0^\beta d\tau_1 d\tau_2\left(\Sigma G - \mathcal{J}^2\left(\frac{2^{q-1}}{q^2}G^{q} + s^2\frac{2^{\tilde{q}-1}}{\tilde{q}^2}G^{\tilde{q}}\right)\right) ~,
\end{equation}
from which we get a set of deformed Schwinger-Dyson equations
\begin{eqnarray}\label{deformed SD equations}
   G^{-1}(\tau_1,\tau_2) &=&  \delta(\tau_1-\tau_2)\partial_{\tau_2}-\Sigma(\tau_1,\tau_2) ~,\\
    \Sigma(\tau_1,\tau_2) &=& \mathcal{J}^2\left(\frac{2^{q-1}}{q}  G(\tau_1,\tau_2)^{q-1} + s^{2} \frac{2^{\tilde{q}-1}}{\tilde{q}} G(\tau_1,\tau_2)^{\tilde{q}-1} \right) ~. \label{deformed SD equations ii}
\end{eqnarray}
As for the case of the single SYK model, these deformed models also simplify in the large $q$, $\tilde{q}$ limit. In particular, they exhibit solvable properties \cite{Jiang:2019pam, Anninos:2020cwo} when both $q$ and $\tilde{q}$ are taken to infinity while keeping their ratio, $q/\tilde{q}$ finite and fixed. From now onwards, we will call this ratio $n \equiv q/\tilde{q} \geq 1$. In this limit, we can again expand the two-point function as in (\ref{large q two point}) and obtain that the Schwinger-Dyson equations simplify to the following ordinary differential equation
\begin{equation}
\label{large q diff equation general ratio}
    \partial_{\tau}^{2}g(\tau) = 2 n s^2\mathcal{J}^2e^{g(\tau)/n} + 2\mathcal{J}^2e^{g(\tau)}~.
\end{equation}
To leading order in the large $q$ and $\tilde{q}$ expansion, the free energy of the deformed model reduces to
\begin{equation}
\label{large q free energy}
\frac{\beta F}{N} = -S^{\mathrm{free}}_0 - \frac{\beta}{8 q^{2}} \int_{0}^{\beta} d \tau\left[\frac{1}{2}\left(\partial_{\tau} g(\tau)\right)^{2}+\mathcal{J}^{2}\left(2 n^2 s^{2} e^{g(\tau)/n}+2e^{g(\tau)}\right)\right]~.
\end{equation}

\subsection{An analytically solvable deformation}

When $n=2$, the differential equation \eqref{large q diff equation general ratio} reduces to %
\begin{equation}
\label{n=2 diff equation general ratio}
    \partial_{\tau}^{2}g(\tau) = 4 s^2\mathcal{J}^2e^{g(\tau)/2} + 2\mathcal{J}^2e^{g(\tau)}~.
\end{equation}
Provided that $g(0) = g(\beta) = 0$, we obtain the following two-point function,\footnote{This solution is slightly different to the one appearing in \cite{Anninos:2020cwo}. The reason is that the model studied there has fermions with two different flavours and so effectively, the number $q$ of fermionic interactions in each term of the Hamiltonian was twice the one considered here. It is possible to recover the solution in \cite{Anninos:2020cwo} by simply taking $g(\tau) \to 2g(\tau)$.}
\begin{equation}
e^{g(\tau)} = \frac{4 \nu ^4}{\left(\sqrt{(\beta {\cal{J}})^2 \nu ^2+ s^4 (\beta  {\cal{J}})^4} \cos (\nu  ({2 \tau}/{\beta}-1))+ s^2 (\beta {\cal{J}})^2\right)^2} \,\,\,\, , \,\,\,\, \cos \nu = \frac{2 \nu ^2- s^2 (\beta {\cal{J}})^2}{\sqrt{(\beta {\cal{J}})^2 \nu ^2+ s^4 (\beta {\cal{J}})^4}} \,.
\end{equation}
Note that this equation provides a solution for the full RG flow for all values of $\beta \mathcal{J}$ and $s$, even in the strongly coupled regime of the theory. We can obtain the free energy by substituting this solution into the on-shell action (\ref{large q free energy}) with $n=2$.  A key observation \cite{Anninos:2020cwo} is that at low temperatures $\beta \mathcal{J} \gg 1$ and small $s \ll 1$, there are two different regimes where the entropy is linear in the temperature. Both regimes can be described analytically. We refer to the regimes  as the deep IR and the intermediate IR regimes, given that thet both appear in the infrared sector of the theory. 

First, let us consider the very small temperature regime, $\beta\mathcal{J}\gg1/s^{2}$, which we refer to as the deep IR regime. In this regime, the entropy is given by \cite{Anninos:2020cwo}
\begin{equation}
\label{deformed deep IR entropy}
  \text { Deep IR: } \quad \frac{S}{N} = \left(S^{\mathrm{free}}_0 -\frac{\pi^{2}}{4\tilde{q}^{2}}\right)+\bar{\aleph}\frac{\pi^{2}}{\tilde{q}^{2}} \frac{1}{s \beta \mathcal{J}}+\cdots  \, ,
\end{equation}
where
\begin{equation}
\label{aleph bar}
    \bar{\aleph} = \frac{\sqrt{1+4s^2}}{2s}~.
\end{equation}
We can compare (\ref{deformed deep IR entropy}) to the IR behaviour of the single SYK model, $s H_{\tilde{q}}$, which is (\ref{entropy large q}) with $q \to \tilde{q}$ and $\mathcal{J} \to s \mathcal{J}$.
While the zero temperature entropy is unchanged, the deformed model changes dramatically the coefficient of the entropy that is linear in the temperature. This is parameterised by the constant $\bar{\aleph}$. Note that in the limit $s \to \infty$, $\bar{\aleph} \to 1$ and we recover the single SYK result, as expected.

We can also study an intermediate regime in which $1 \ll \beta \mathcal{J} \ll 1/s^{2} $. Given that $\beta \mathcal{J} \gg 1$, we are still in the infrared, so we call this regime intermediate IR. The leading order thermodynamics can also be computed analytically here obtaining that the entropy is given by
\begin{equation}
\label{Intermediate IR entropy}
    \text { Intermediate IR: } \quad \frac{S}{N}=\left(S^{\mathrm{free}}_0-\frac{\pi^{2}}{4q^2}\right) +\frac{\pi^{2}}{q^2}\frac{1}{\beta \mathcal{J}}  +\cdots \,.
\end{equation}
Note that, to leading order, this entropy is independent of $s$ and corresponds to a single SYK Hamiltonian with a $q$ fermion interaction. The first deviation from the linear behaviour will depend on $s$ and is studied in section \ref{sec_int_IR}. 
\newline\newline
In the remainder of the paper we discuss different properties of these deformed models away from this solvable limit.  seen as coming from a Schwarzian action. 

\section{Thermodynamics of deformed SYK} \label{sec_deep_IR}

In this section we analyse the deformed models (\ref{deformed Hamiltonian}) for  general values of $n = q/\tilde{q}$, both at finite and large $q$. An emphasis  is placed on the deep IR behaviour of the deformed model, given by $\beta\mathcal{J}\gg1/s^{2}$. When $n\neq 2$, we must resort to a combination of analytical and numerical techniques to compute thermodynamic quantities. We begin by analysing the large $q$ limit. We compute the large $q$ entropy at low temperatures, from which we can numerically extract the coefficient, $\bar{\aleph} (s, n)$, of the linear-in-temperature part of the entropy, for various values of $n$. We conjecture that a similar structure for the entropy holds for finite values of $q$ and check it against numerical data for $n=2,3,4$, and different finite values of $q$, finding good agreement. We also provide evidence for the existence of models with two near-conformal regimes at both large and finite $q$, characterised by two linear-in-temperature regimes for the entropy. Finally, we uncover a novel analytically tractable window for $n=1+\varepsilon$, with $\varepsilon$ small. 

\subsection{Large $q$}
\label{Large q limit of the deformed SYK with general n}
We start by computing $\bar{\aleph}(s, n)$ numerically for general $n$, in the large $q$ limit. To do so, we need to solve equation (\ref{large q diff equation general ratio}), with boundary conditions $g(0) = g(\beta) = 0$. Given a numerical solution $g(\tau)$, we can compute the free energy following equation (\ref{large q free energy}). The entropy then can be obtained using (\ref{S from F}). Instead of computing the thermodynamic derivative numerically we use that $\beta\partial_{\beta} = \mathcal{J}\partial_{\mathcal{J}}$ \cite{Maldacena:2016hyu} to compute the entropy directly as
\begin{equation}
\label{numerical entropy}
 \frac{S}{N} = S^{\mathrm{free}}_0 + \frac{\beta}{8 n^2 \tilde{q}^2} \int_{0}^{\beta} d \tau\left[\frac{1}{2}\left(\partial_{\tau} g(\tau)\right)^{2}-\mathcal{J}^{2}\left(2 n^2 s^{2} e^{g(\tau)/n}+2 e^{g(\tau)}\right)\right]~.
\end{equation}
In the deep IR, it is more convenient to parameterise formulas in terms of $\tilde{q}$ instead of $q$, as $H_{\tilde{q}}$ is the dominating term in the Hamiltonian in this regime. Our numerical results confirm that at low enough temperatures, $\beta\mathcal{J}\gg1/s^{2}$, the entropy is linear in the temperature, taking the form 
\begin{equation} \label{entropy ansatz}
\frac{S}{N} = \left(S^{\mathrm{free}}_0 + S_{0}(s, n)\right)+\bar{\aleph} (s, n) \frac{\pi^{2}}{\tilde{q}^{2}} \frac{1}{s \beta \mathcal{J}}+\cdots,
\end{equation}
where now  $\bar{\aleph} (s, n)$ can in general depend on $s$ and $n$, but is independent of $\beta \mathcal{J}$. The zero temperature entropy is shifted by a factor $S_0(s, n)$ that may also generally depend on $s$ and $n$. 

\

\begin{figure}[H]
    \centering
    \vspace{0mm}
    \includegraphics[width=0.5\columnwidth]{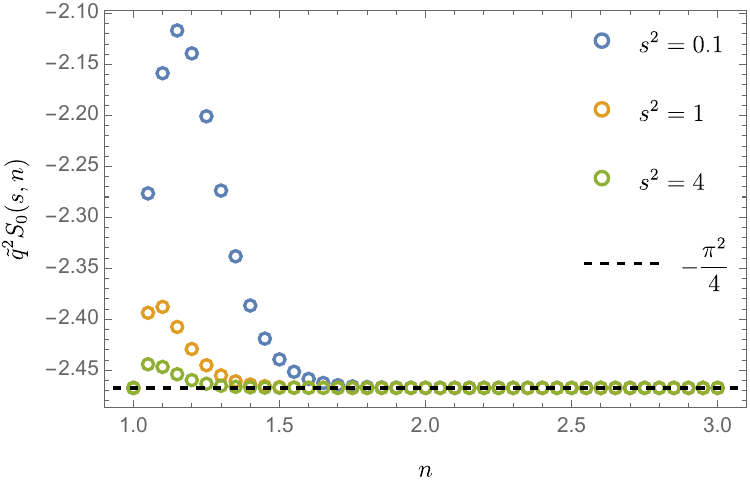}
    \caption{The zero temperature entropy $\tilde{q}^{2}S_{0}(s, n)$ as a function of $n$, for $s^{2}=0.1, 1, 4$. The circles are numerical computations for different values of $s^2$, while the dashed black line indicates the analytic value of $\tilde{q}^{2}S_{0}(s, n)$ for a single SYK model.}\label{fig:zero_point}
\end{figure}

\begin{figure}[H]	
    \centering
    \vspace{0mm}
    \includegraphics[width=0.5\columnwidth]{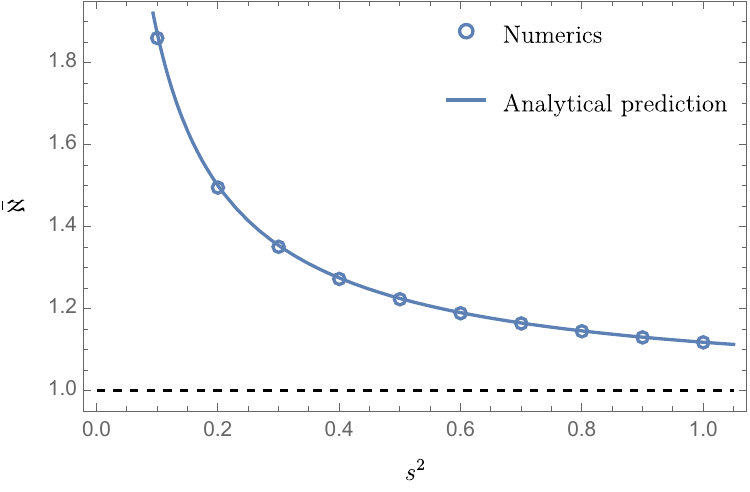} \caption{$\bar{\aleph}$ as a function of $s^2{}$ for the deformed SYK model in the large $q$ limit with $n=2$. The circles are numerical computations while the blue solid curve shows the analytic result in (\ref{aleph bar}), for comparison. At large $s$, we expect the numerics to tend towards the black dashed line at $\bar{\aleph}=1$.}\label{fig:large_q_C_deformed}
\end{figure}

\begin{figure}[H]
    \centering
    \vspace{0mm}
    \includegraphics[width=0.5\columnwidth]{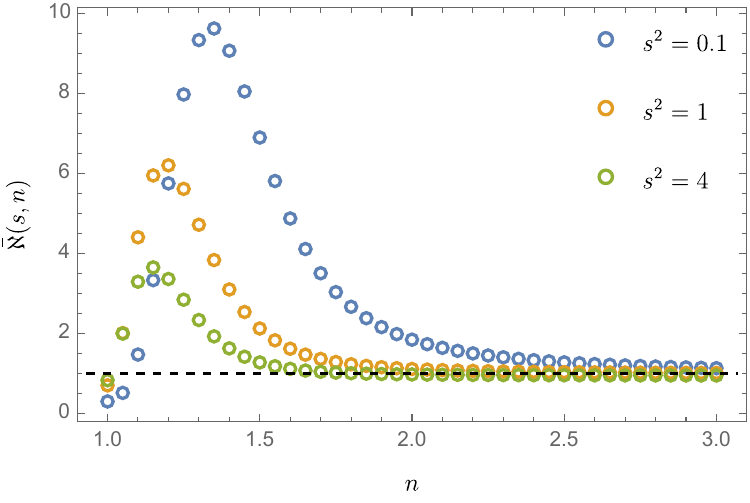}
    \caption{$\bar{\aleph}(s,n)$ as a function of $n$ for $s^{2}=0.1, 1, 4$. The circles are numerical computations. For large $n$, $\bar{\aleph}(s,n)$ tends towards the expected value of $\bar{\aleph}(s,n)=1$ shown in a dashed black line.}\label{fig:aleph_bar}
\end{figure}

\noindent \textbf{Zero temperature entropy.} We can find $\tilde{q}^{2}S_0(s, n)$ numerically by performing a linear fit of $\tilde{q}^2\beta\mathcal{J}\left(S/N -S^{\mathrm{free}}_0 \right)$ as a function of $\beta\mathcal{J}$ for large values of $\beta\mathcal{J}$. In figure \ref{fig:zero_point}, we show results for $s^{2}=0.1, 1, 4$ with $1\leq n \leq 3$, using values of $\beta\mathcal{J}$ between 2000 and 3000 for the linear fit. We find that for $n\geq2$, the shift in the zero temperature entropy is given by $\tilde{q}^{2}S_0(s, n) =  -{\pi^{2}}/{4}$, the same as that of a single SYK model with Hamiltonian $sH_{\tilde{q}}$. As shown in figure \ref{fig:zero_point}, there are deviations from the single SYK result within the interval $1< n < 2$, but they vanish as $n \to 2$. The $s$ dependence of the entropy at vanishing temperature, as well as the transition at $n=2$, and their potential holographic interpretation, merit a deeper understanding perhaps along the lines of \cite{Affleck:1991tk}. We will return to this in future work. 
$\,$\newline
\noindent \textbf{The deep IR phase at large $q$.}
We numerically compute the entropy $\tilde{q}(S/N - S^{\mathrm{free}}_0)$ at a single low temperature point.\footnote{In figures \ref{fig:large_q_C_deformed} and  \ref{fig:aleph_bar} we present numerical results for $\beta\mathcal{J} = 3000$. We have also performed this procedure for other values of $\beta\mathcal{J}$ between $2000$ and $3000$ allowing us to test the postulated $\beta$-dependence of (\ref{entropy ansatz}).} Subtracting the previously obtained values for $\tilde{q}^{2}S_0(s, n)$ from this, the leading contribution to the difference is a term that is proportional to $(\beta \mathcal{J})^{-1}$, from which we can numerically extract $\bar{\aleph} (s, n)$ in \eqref{entropy ansatz}. For $n=2$, there is an analytic answer for $\bar{\aleph}$ given by (\ref{aleph bar}). We use this as a consistency check of our numerical procedure. In figure \ref{fig:large_q_C_deformed}, we show agreement between our numerical algorithm and the analytic result for $n=2$.
For $n \neq 2$, there are no known analytic solutions. However, we do expect a certain behaviour of $\bar{\aleph} (s, n)$ in a variety of limits. Namely,
\begin{itemize}
\item[1.] For $s\to\infty$ and fixed $n$, we expect the leading entropy to be that of a single SYK model with Hamiltonian $sH_{\tilde{q}}$ and so, $\bar{\aleph} (s\to \infty, n) \to 1$ in this limit.

\item[2.] At fixed $s$ but $n\to\infty$, we also expect $\bar{\aleph} (s, n\to \infty) \to 1$. To see this, note that $n \to \infty$ implies $q \to \infty$ with $\tilde{q}$ finite.  The contribution to the free energy from $H_q$ is given by $2^{q-1} q^{-2} G^{q}$, see \eqref{deformed bilocal action}. Given that $|G(\tau)| \leq 1/2$, if we take $q$ to infinity this contribution is negligible and only the terms with $\tilde{q}$ will contribute. Thus, $\bar{\aleph} (s, n\to \infty) \to 1$.

\item[3.] When $n=1$, the theory is equivalent to a single SYK with Hamiltonian $\sqrt{1+s^2}H_{q}$. We therefore expect that
\begin{equation}\label{aleph bar n=1}
    \bar{\aleph}(s, n=1) = \frac{s}{\sqrt{1+s^2}}~.
\end{equation}

\item[4.] Finally, as discussed, when $n=2$, we know analytically that
\begin{equation}
    \bar{\aleph}(s, n=2) = \frac{\sqrt{1+4s^2}}{2s}~.
\end{equation}
\end{itemize}
In figure \ref{fig:aleph_bar} we plot numerical values of $\bar{\aleph}(s, n)$ as a function of $n$ for different values of $s^2$. We see that the numerical results behave as expected in the limits mentioned above. When $n=1$ and $n=2$, the numerical values agree with the analytically known values. We also observe that as $s^2$ grows deviations from $\bar{\aleph}(s, n) = 1$  decrease for all values of $n$, consistent with the expectation that when $s$ becomes large $\bar{\aleph}(s, n) \to 1$. Furthermore, as $n$ becomes large we see that $\bar{\aleph} (s,n) \to 1$, as expected. 

We also notice an interesting behaviour of $\bar{\aleph} (s,n)$ between $n=1$ and $n=2$, characterised by a peak whose position depends on $s$. Following the analytic arguments on section \ref{Schwarzian for the deep IR}, we expect the peak to move towards $n=3/2$, as $s$ becomes smaller. Though we were unable to find a general analytic form for $\bar{\aleph}(s, n)$, the numerical results suggest that, at least at small $s$ and $n \geq 2$, the empirical formula
\begin{equation}
\label{aleph for small s empirical main}
    \bar{\aleph}(s,n) \approx \frac{a(n)}{s^{4/n^2}}  \,,
\end{equation}
holds with $1/2 \leq a(n) \leq 1$. More details on this are provided in Appendix \ref{section: Towards an analytical form for alephbar}.

\begin{figure}[H]
        \centering
         \subfigure[$n=3$, full RG flow]{
                \includegraphics[height=4.44cm]{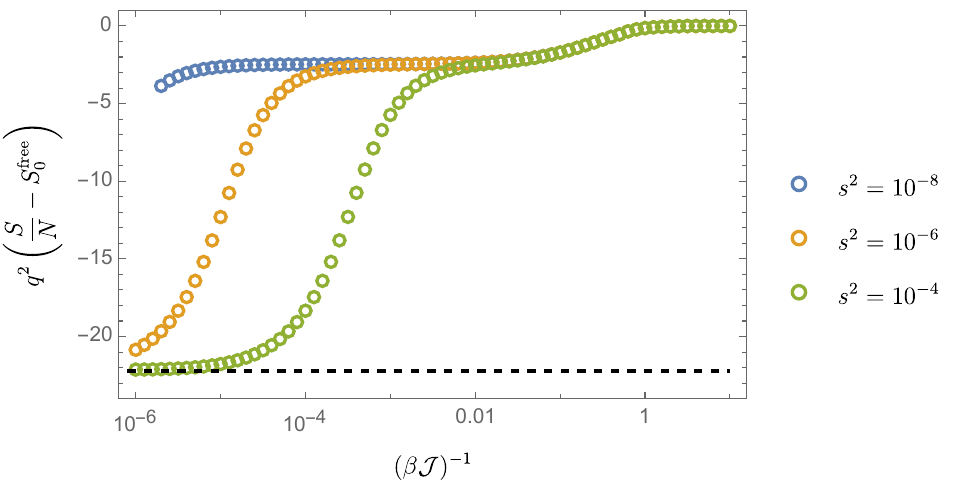}\label{fig:n_3_full}}  \quad
        \subfigure[$n=3$, intermediate IR]{
                \includegraphics[height=4.44cm]{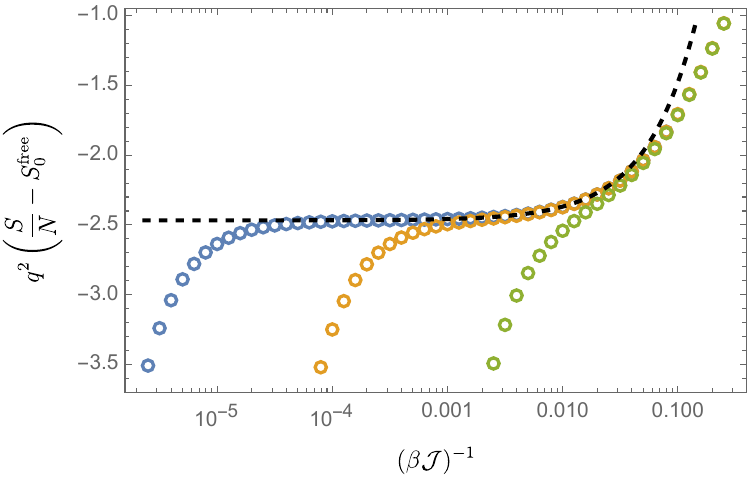} \label{fig:n_3_intermediate}} 
                 \caption{The entropy as a function of temperature (in logarithmic scale) for the deformed SYK model at large $N$ and large $q$ with $n=3$. In \ref{fig:n_3_full} we plot the full RG flow accessible to our numerics. The dashed line gives the expected zero temperature entropy (see figure \ref{fig:zero_point}, noting that in this case $q=3\tilde{q}$). In \ref{fig:n_3_intermediate} we zoom into the intermediate IR regime. The dashed line gives the expected analytic form \eqref{Intermediate IR entropy}.}\label{fig:largeqRG}
\end{figure}

\noindent \textbf{The intermediate IR phase at large $q$.} For large values of $q$ and $n \geq 2$, the RG flow at small enough $s$ develops two near-fixed points. At finite temperature this is revealed by the presence of two linear-in-temperature regimes for the entropy. We find that, just as in the $n=2$ case, the leading order entropy in the intermediate IR regime is given by \eqref{Intermediate IR entropy}. An example of this behaviour, for $n=3$, is given in figure \ref{fig:largeqRG}. 
In sections \ref{Large N, finite q deformed} and \ref{CPT} we provide evidence for the existence of a near-fixed point at finite $q$. Moreover, in section \ref{CPT} we present evidence of the intermediate fixed point for $1<n<2$ at large $q$. A systematic analysis of the behaviour in the proximity of the two near-fixed points is discussed in section \ref{sec_int_IR}.

\subsection{Finite $q$}
\label{Large N, finite q deformed}

Given the results in the large $q$ limit, we now analyse the case of finite $q$. This is numerically more involved than the previous case, as the Schwinger-Dyson equations no longer reduce to an ordinary differential equation. Instead, we need to solve the Schwinger-Dyson equations (\ref{deformed SD equations}) and (\ref{deformed SD equations ii}) numerically. This set of equations is amenable to numerical computations using a recursive algorithm and the fast Fourier transform. In Appendix \ref{appendix: numerical} we outline the details of this procedure, which is analogous to the one described in Appendix G of \cite{Maldacena:2016hyu} for the single SYK model. The simplest deformed model at finite $q$ has $q= 4$ and $\tilde{q} = 2$, which is first studied in \cite{Garcia-Garcia:2017bkg}. In the present work, we extend this analysis to include smaller values of $s^2$, allowing us to observe two different near-conformal regimes. We also present results for a more general class of models with different values of $q$ and $\tilde{q}$.
\newline\newline
\noindent \textbf{The deep IR phase at finite $q$.} We start by focussing on the form of the entropy in the deep IR limit. We have numerical access to this regime provided $s$ is not very small. For a single SYK model with $\tilde{q}$ and coupling $s \mathcal{J}$, the entropy in the limit $\beta \mathcal{J}\gg1/s$ is given by
\begin{equation}
    \frac{S}{N} = 
     \left(S^{\mathrm{free}}_0 - \int_{0}^{1/\tilde{q}} dx\; \pi\left(\frac{1}{2}-x\right) \tan \pi x\right) +\frac{ 4\pi^{2}\alpha(\tilde{q})}{s \beta\mathcal{J}}+\cdots ~,
\end{equation}
where $\alpha (\tilde{q})$ is the same (numerical) coefficient that appeared in the Schwarzian action in \eqref{Scwharzian action} (see Appendix \ref{app: alpha} for more detail). 

\begin{figure}[H]
        \centering
         \subfigure[$q=4, \tilde{q}=2$]{
                \includegraphics[height=4.36cm]{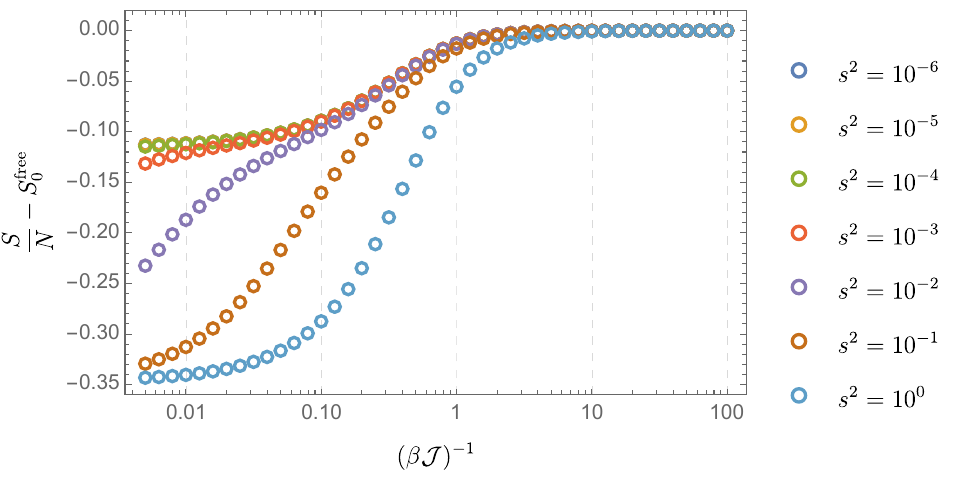}\label{fig:4_2_finiteq}}  \quad
        \subfigure[$q=6, \tilde{q}=2$]{
                \includegraphics[height=4.36cm]{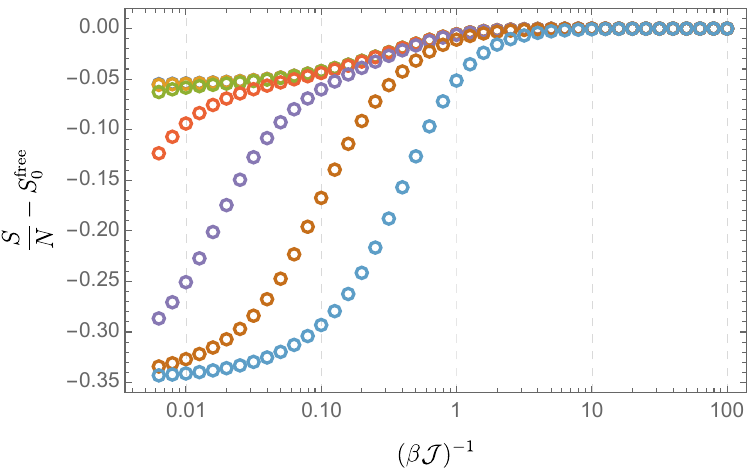} \label{fig:6_2_finiteq}} 
                 \caption{The entropy as a function of temperature (in logarithmic scale) for the deformed SYK model at large $N$ and finite $q$. Different colours correspond to different values of $s^2$. Circles correspond to numerical computations.}\label{fig:entropy_modified_SYK}
\end{figure}

Moving to the case of the deformed Hamiltonian, we first discuss the case of $n=2$. In section \ref{Large q limit of the deformed SYK with general n}, we found that for $n\geq 2$ the zero temperature entropy of the deformed model was the same as that of a single SYK. Assuming this is the case even at finite $q$, we propose that the entropy in the deformed theory should be generalised to 
\begin{equation}
\label{finite q specific heat ansatz}
    \frac{S}{N} = 
     \left(S^{\mathrm{free}}_0 - \int_{0}^{1/\tilde{q}} dx\; \pi\left(\frac{1}{2}-x\right) \tan \pi x\right) + \bar{\aleph}\,  \frac{4\pi^{2} \alpha(\tilde{q}) }{s \beta \mathcal{J}}+\cdots ~.
\end{equation}
Namely, the zero temperature entropy remains the same and the linear-in-temperature term gets an extra coefficient of $\bar{\aleph}$ -- as defined in (\ref{aleph bar}) -- with respect to the single SYK theory. We numerically find that for large $s$ and low temperatures, $(S/N-S^{\mathrm{free}}_{0})$ approaches the predicted value of $-0.346$ obtained from setting $\tilde{q} = 2$ in \eqref{finite q specific heat ansatz} (see for example figure \ref{fig:entropy_modified_SYK}).

\begin{figure}[H]
        \centering
         \subfigure[$q=4, \tilde{q}=2$]{
                \includegraphics[scale=0.5]{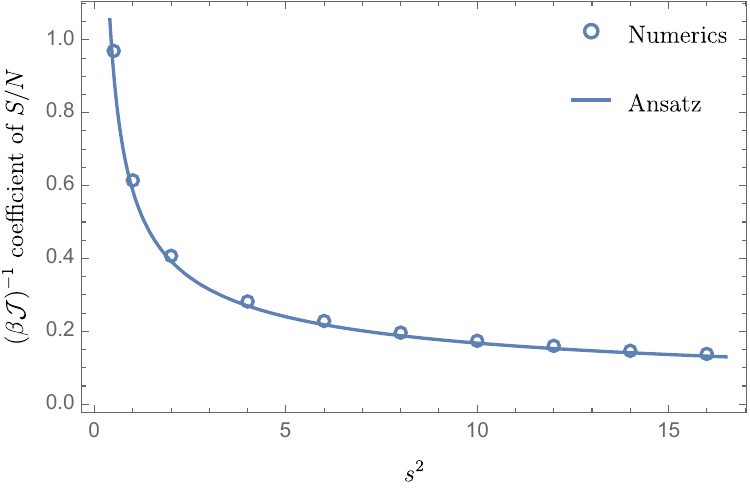}\label{q_2_4}}  \quad\quad
        \subfigure[$q=8, \tilde{q}=4$]{
                \includegraphics[scale=0.5]{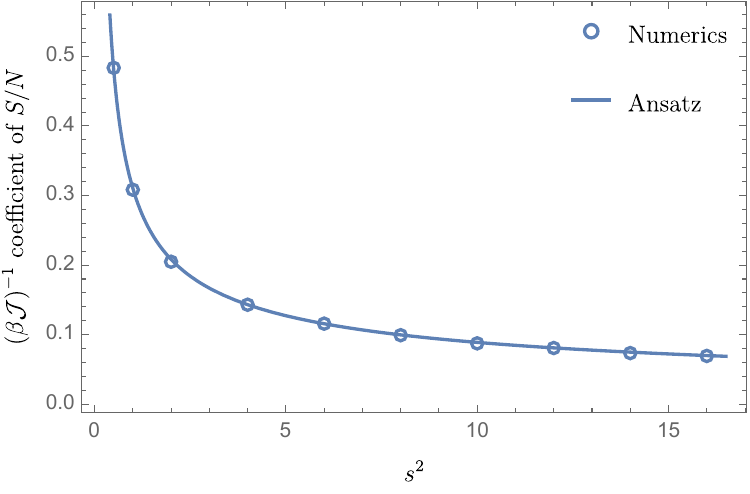} \label{q_4_8}}  \quad\quad
       \subfigure[$q=12, \tilde{q}=6$]{
                \includegraphics[scale=0.5]{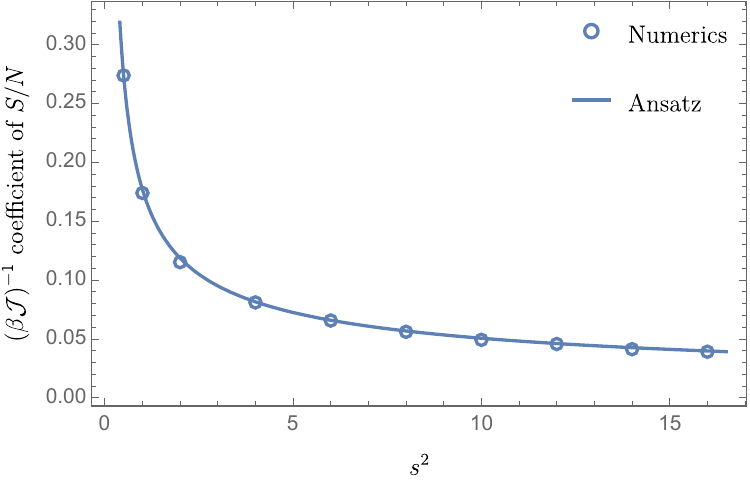} \label{q_6_12}}  \quad\quad
       \subfigure[$q=16, \tilde{q}=8$]{
                \includegraphics[scale=0.5]{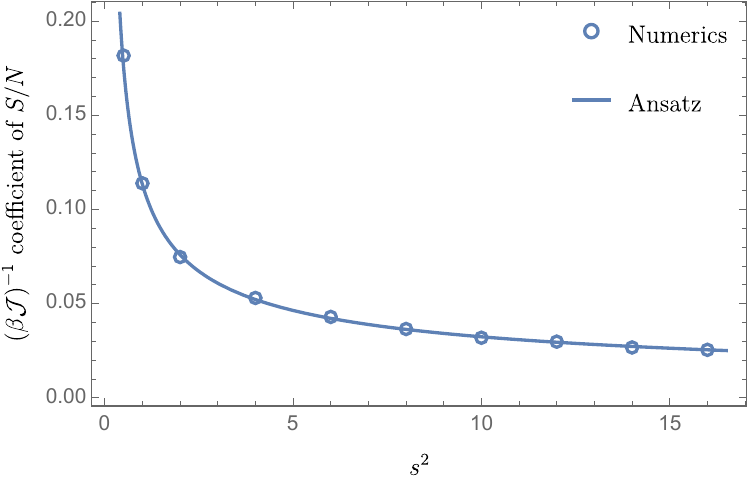} \label{q_8_16}}                         
                \caption{The linear-in-temperature coefficient of the entropy as a function of $s^2$, in the deformed SYK with $n=2$ for finite $q$ and $\tilde{q}$. The circles correspond to numerical computations while the blue solid curve corresponds to \eqref{finite q specific heat ansatz}, conjectured from the large $q$ limit behaviour.}\label{fig:finite_q_C_deformed}
\end{figure}

To test the linear-in-temperature coefficient, we compute the entropy at a single low temperature point and subtract the zero temperature entropy. In figure \ref{fig:finite_q_C_deformed}, we show the numerical results for the coefficient and compare to the analytic prediction, as in (\ref{finite q specific heat ansatz}), for different values of $q$ and $\tilde{q}$, with fixed $n=2$. To compute the predicted coefficient, we use values of $\alpha(\tilde{q})$ obtained from the Pad\'e approximation as described in Appendix \ref{app: alpha} and the analytic value of $\bar{\aleph}$ for $n=2$ in the large $q$ limit. We find remarkable agreement, suggesting the possibility of using large $q$ (analytical) results to extract finite $q$ information.

\begin{figure}[H]
        \centering
         \subfigure[$q=12, \tilde{q}=4$]{
                \includegraphics[scale=0.55]{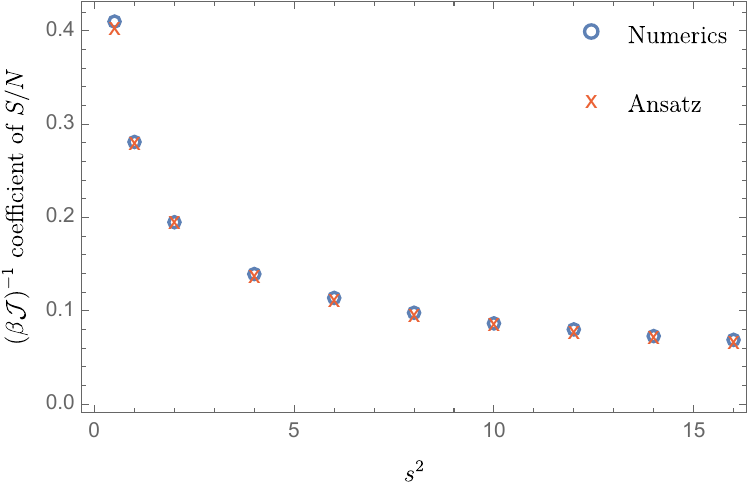}\label{q_4_12}}  \quad\quad
       \subfigure[$q=16, \tilde{q}=4$]{
                \includegraphics[scale=0.55]{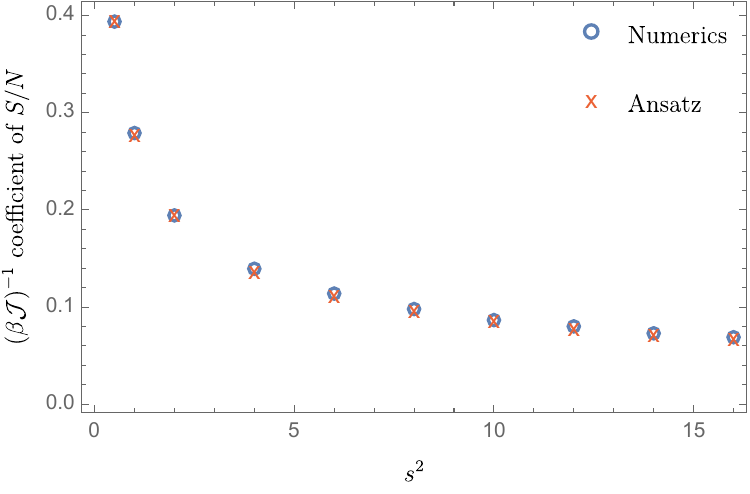} \label{q_4_16}}                         
                \caption{The linear-in-temperature coefficient of the entropy as a function of $s^2$, in the deformed SYK for $n=3,4$ with finite $q$ and $\tilde{q}$. The circles correspond to numerical computations while the crosses correspond to \eqref{general n ansatz} with the value of $\bar{\aleph}(s, n)$ obtained numerically in the large $q,\tilde{q}$ limit.}\label{fig:finite_q_n_3_4}
\end{figure}

The results for $n=2$ hint towards the possibility of generalising the form of the low temperature entropy even away from the $n=2$ point. In fact, following the results at large $q$, we propose that the only change in the form of the entropy (\ref{finite q specific heat ansatz}) for $n > 2$ is to take $\bar{\aleph} \to \bar{\aleph} (s,n)$, where $\bar{\aleph} (s,n)$ is the coefficient obtained numerically in the large $q$ limit, see figure \ref{fig:aleph_bar}. Note that for $1<n<2$ we would also expect a change in the zero point temperature, as is seen at large $q$. The proposal, then, is that, at finite $q$, for $n\geq2$, the low temperature entropy takes the form,
\begin{equation}
\label{general n ansatz}
    \frac{S}{N} = \left(S^{\mathrm{free}}_0 - \int_{0}^{1/\tilde{q}} dx\; \pi\left(\frac{1}{2}-x\right)\tan \pi x\right) + \bar{\aleph}(s,n) \frac{4\pi^{2}\alpha(\tilde{q}) }{s\beta \mathcal{J}}~.
\end{equation}
We test this conjecture for $n=3$ and $n=4$ by numerically computing the entropy for for $q=12$, $\tilde{q}=4$ and $q=16$, $\tilde{q}=4$ respectively. As before we use a single low temperature point and subtract the zero temperature entropy to isolate the linear-in-temperature coefficient. To compute the predicted linear-in-temperature coefficient, as in \eqref{general n ansatz}, we again use values of $\alpha(\tilde{q})$ from the Pad\'e approximant described in Appendix \ref{app: alpha} but now use values of $\bar{\aleph}(s,n)$ obtained numerically at large $q$. The results are shown in figure \ref{fig:finite_q_n_3_4}, demonstrating strong evidence.


\noindent \textbf{The intermediate IR phase at finite $q$.}  We now provide evidence that even at finite $q$, the RG flow at small enough $s$ develops two near-conformal regimes. We consider the cases of $n=2$ with $q= 4$ and $\tilde{q} = 2$ and $n=3$, with  $q= 6$ and $\tilde{q} = 2$. 
In figure \ref{fig:entropy_modified_SYK}, we plot entropy as a function of $(\beta \mathcal{J})^{-1}$ for different values of the coupling $s^2$, from $s^2 =1$ to $s^2 = 10^{-6}$, for both models. In each case, at large temperatures, all the curves approximate the entropy of the free fermions. As we move towards the IR, and similar to what happens at large $q$, there are two clearly different behaviours depending on the value of $s^2$. When $s^2 \sim 1$, the entropy goes directly into the deep IR phase. When $s^2 \ll 1$, there is a different intermediate IR phase appearing with a linear-in-temperature regime. It is natural to suspect that at even lower temperatures, these theories will also end up flowing into the deep IR phase. However, the numerical techniques employed are only powerful enough to reach $(\beta \mathcal{J})^{-1} \gtrsim 10^{-3}$. This does not permit us to compute a full RG flow exhibiting both the intermediate and the deep IR phase. Implementing an algorithm based on spectral methods might provide an efficient way of reaching even lower temperatures of order at least $(\beta \mathcal{J})^{-1} \sim 10^{-4}$ \cite{Cruz:2022uic}. We leave such an approach for future work.

\subsection{Large $q$ with $n=1+\varepsilon$}
\label{sec: Large q, n=1+epsilon}
To finish this section we discuss a novel analytically tractable RG flow at large $q$, for $n=1+\varepsilon$, with $\varepsilon$ a small positive number.

We first discuss the leading order solution $g_0(\tau)$ with $n=1$. At the level of the effective action \eqref{deformed bilocal action}, the deformed model with $n=1$ is equivalent to a single SYK model with random couplings averaged over a Gaussian distribution with a variance proportional to $\mathcal{J}^{2}(1+s^{2})$. In fact, at large $q$, the differential equation \eqref{large q diff equation general ratio} for $n=1$, becomes
\begin{equation} \label{g order zero}
    \partial_{\tau}^{2}g_0(\tau) = 2\mathcal{J}^{2}(1+s^2) e^{g_0(\tau)}~,
\end{equation}
which after imposing thermal boundary conditions, $g_0(0) = g_0(\beta) = 0$, is solved by
\begin{equation} \label{finite temp g0}
e^{g_0(\tau)} = \frac{\cos^2 \nu}{\cos^2 \left( 2 \nu \left( \frac{1}{2} - \frac{|\tau|}{\beta} \right)\right)} \, , \quad\quad \beta \mathcal{J}  = \frac{2 \nu}{\sqrt{1+s^2} \cos \nu} \,.
\end{equation}
We now consider $n= 1 + \varepsilon$, perturbatively in $\varepsilon$. We can expand $g(\tau)$ as 
\begin{equation}
\label{epsilon expansion of g}
    g(\tau) = g_{0}(\tau) + \varepsilon g_{1}(\tau) + \mathcal{O}(\varepsilon^2).
\end{equation}
Substituting this into the differential equation \eqref{large q diff equation general ratio}, we find, at leading order in $\varepsilon$, a differential equation for $g_1 (\tau)$
\begin{equation}
\label{g1 diff eqn}
    \partial_\tau^2 g_1(\tau) = 2e^{g_0(\tau)}  \mathcal{J}^{2}\left((1-g_0(\tau)) s^2+ g_1(\tau) \left(1+s^2\right)\right)~.
\end{equation}
It is straightforward to show that
\begin{equation}
 g_1(\tau) = \frac{s^2}{1+s^2}g_0(\tau) \,,
\end{equation}
is the solution to \eqref{g1 diff eqn} with boundary conditions $g_1(0) = g_1(\beta) = 0$. To see this, note that if we plug this expression for $g_1(\tau)$ in (\ref{g1 diff eqn}), we get that
\begin{equation}
    \partial_\tau^2 g_0 (\tau) = 2\mathcal{J}^{2}(1+s^2)e^{g_0 (\tau)}~,
\end{equation}
which is exactly (\ref{g order zero}), so it is satisfied by $g_0(\tau)$. Next, we consider the corrections to the free energy coming from this deformation. Expanding \eqref{large q free energy} to leading order in $\varepsilon$ we obtain
\begin{equation}\label{free energy near 1}
  \left.  \frac{\beta F}{N} \right|_{n=1+\varepsilon} =   -S^{\mathrm{free}}_0 +\frac{\nu(\nu-2\tan{\nu})}{\tilde{q}^{2}}-\frac{2 \nu (\nu -2 \tan \nu)}{1+s^2} \frac{\varepsilon}{\tilde{q}^2} + \mathcal{O}(\varepsilon^{2})~.
\end{equation}
Using \eqref{S from F} we find the entropy to leading order in $\varepsilon$ is given by
\begin{equation}\label{entropy nu near n_1}
 \left. \frac{S}{N} \right|_{n=1+\varepsilon}= S^{\mathrm{free}}_0 -\frac{\nu ^2}{\tilde{q}^2}  + \frac{ 2  \nu ^2}{1+s^2} \frac{\varepsilon}{ \tilde{q}^2}  + \mathcal{O}(\varepsilon^{2})~.
\end{equation}
This can be used to find the entropy as a function of temperature for the full RG flow. Though we do not observe an intermediate IR at this order in $\varepsilon$, we are able to access some interesting features of the deep IR. Expanding \eqref{entropy nu near n_1} in powers of $(\beta \mathcal{J})^{-1}$ we find the correction to the entropy,
\begin{equation}
\label{entropy correction}
    \left. \frac{S}{N} \right|_{n=1+\varepsilon} = \left. \frac{S}{N} \right|_{n=1} + \left( \frac{\pi ^2}{2 \left(1+s^2\right)}-\frac{2\pi ^2}{\left(1+s^2\right)^{3/2}}\frac{1}{\beta  \mathcal{J}}  +\mathcal{O}\left(\beta\mathcal{J}\right)^{-2}\right) \frac{\varepsilon}{\tilde{q}^{2}}  + \mathcal{O}(\varepsilon^{2})~,
\end{equation}
where the entropy at low temperatures for $n=1$ is given by equation (\ref{entropy large q}) with $\mathcal{J} \to \sqrt{1+s^2} \mathcal{J}$ and $q \to \tilde{q}$. Equation (\ref{entropy correction}) provides two predictions that can be tested against numerical computations. We study these next.

\

\begin{figure}[H]
    \centering
    \vspace{0mm}
    \includegraphics[width=0.45\columnwidth]{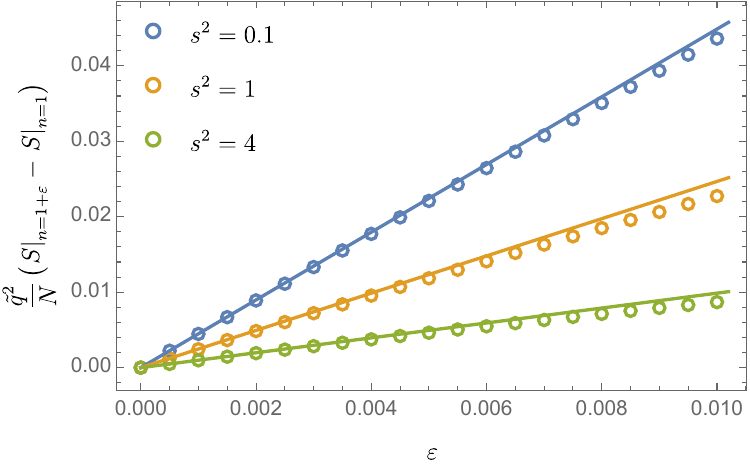}
    \caption{The difference in the zero temperature entropy between the large $q, \tilde{q}$ model with $n=1+\varepsilon$ and $n=1$, as function of small $\varepsilon$, for different values of $s^2$. The circles correspond to numerical computations at $\beta \mathcal{J} =2000$, while the solid lines are the analytic prediction from \eqref{zero point correction}. For small enough $\varepsilon$, both overlap.}\label{fig:zero_point_correction}
\end{figure}

\noindent \textbf{Zero temperature entropy.}
Note that the correction to the zero temperature entropy at large $\tilde{q}$ is given by
\begin{equation}
\label{zero point correction}
\lim_{\beta\mathcal{J} \to \infty} \left. \frac{\tilde{q}^2 S (\beta \mathcal{J}  )}{N} \right|_{n=1+\varepsilon} - \left. \frac{\tilde{q}^2 S(\beta \mathcal{J} )}{N} \right|_{n=1} =   \frac{\pi^{2}}{2(1+s^{2})} \varepsilon+ \mathcal{O}(\varepsilon^{2})~.
\end{equation}
We can numerically compute the large $q, \tilde{q}$ entropy for $n=1$ and for $n=1+\varepsilon$ at large $\beta\mathcal{J}$ for small values of $\varepsilon$ and compare with the analytic prediction. We show the results for $s^2 = 0.1, 1, 4$ at $\beta \mathcal{J} = 2000$ in figure \ref{fig:zero_point_correction}, showing agreement between the analytical predictions and the numerical computations.

\begin{figure}[H]
    \centering
    \vspace{0mm}
    \includegraphics[width=0.5\columnwidth]{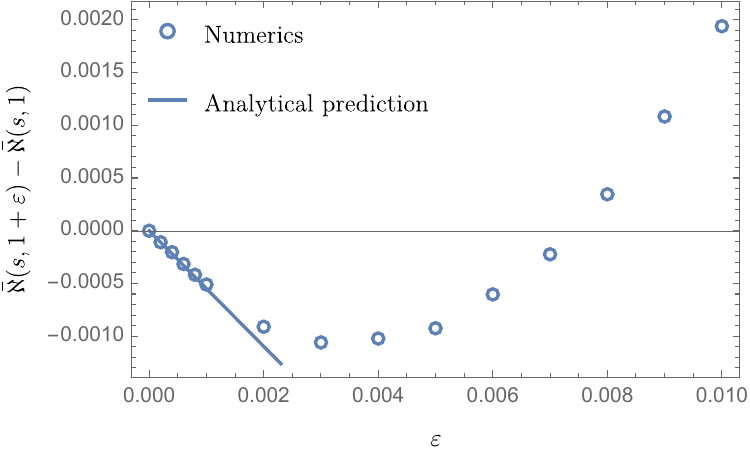}
    \caption{Difference in the values of $\bar{\aleph}(s, n)$ between $n=1+\varepsilon$ and $n=1$, as a function of small values of $\varepsilon$, with $s^2=0.1$. The circles correspond to numerical computations, while the solid blue line is the analytic result from \eqref{aleph bar correction}. Note that they match at small $\varepsilon$, showing that $\bar{\aleph}(s, n)$ initially decreases as $n$ moves away from $n=1$. For larger $\varepsilon$, $\bar{\aleph}(s, n)$ starts increasing again, which agrees with the results shown in figure \ref{fig:aleph_bar}. We do not see the initial decrease in $\bar{\aleph}(s, n)$ in figure \ref{fig:aleph_bar} since the lowest $\varepsilon$ considered there is $\varepsilon = 0.05$, much larger than the values shown in this plot.}\label{fig:aleph_bar_correction}
\end{figure}

\noindent \textbf{Linear-in-temperature entropy.}
We can also find analytically the correction to the linear-in-temperature term in the entropy, and from this the correction $\bar{\aleph}(s, n)$ near $n=1$. From \eqref{entropy correction}, we find
\begin{equation}\label{aleph bar correction}
    \bar{\aleph}(s, 1+\varepsilon) - \bar{\aleph}(s, 1)  = - \varepsilon \frac{2s}{(1+s^2)^{3/2}} + \mathcal{O}(\varepsilon^{2})~,
\end{equation}
where as $\bar{\aleph}(s, 1)$ is given by \eqref{aleph bar n=1}. Note that the expected value of $\bar{\aleph}(s, n)$ is lower than the value for $n=1$. In figure \ref{fig:aleph_bar_correction}, we test the predicted correction in \eqref{aleph bar correction} against numerical computations for $s^2 = 0.1$ and small values of $\varepsilon$, finding remarkable agreement. 

\section{Conformal perturbation theory} \label{sec_int_IR}

In this section we explore thermodynamic contributions to the free energy and entropy of the deformed SYK near each fixed point. We argue that the leading terms in the entropy expansions \eqref{deformed deep IR entropy} and \eqref{Intermediate IR entropy} can be understood as perturbations to the conformal actions of the single SYK models $sH_{\tilde{q}}$ and $H_q$, respectively. In particular, we will argue that in both cases, the leading irrelevant correction to the free energy, which is proportional to the temperature, stems from a Schwarzian action. Moreover, in the intermediate IR regime, the leading relevant correction away from the intermediate fixed point can be understood from a relevant conformal operator in conformal perturbation theory.
\subsection{Schwarzian for the deep IR}
\label{Schwarzian for the deep IR}


In section \ref{Large N, finite q deformed} numerical evidence was presented indicating that the entropy, $S$, for the finite $q$ deformed model in the deep IR takes the low temperature expansion
\begin{equation}
    \frac{S}{N} = 
    \text{const} + \bar{\aleph}(s, n) \frac{4\pi^{2} \alpha(\tilde{q}) }{s \beta \mathcal{J}} +\cdots~.
\end{equation}
The linear-in-temperature part in $S$ is modified from that of an undeformed SYK model with Hamiltonian $sH_{\tilde{q}}$ by $\bar{\aleph}(s, n)$. 

We would like to understand the linear-in-temperature part in $S$  as coming from the leading correction to a conformal piece of the action associated with the SYK Hamiltonian $sH_{\tilde{q}}$ \cite{Rosenhaus:2018dtp}. More explicitly, by taking $\Sigma \to \Sigma + \partial_{\tau}$, we can re-write the $G\Sigma$-action \eqref{deformed bilocal action} as $I = \tilde{I}_{\text{CFT}} + \tilde{I}_{\text{UV}}$ where
\begin{eqnarray}\label{CFT action deep IR}
   \tilde{I}_{\text{CFT}}&=&-\frac{1}{2}\log\det(-\Sigma) + \frac{1}{2}\int_{0}^{\beta}\int_{0}^{\beta} d\tau_1 d\tau_2\left(\Sigma G - s^2\mathcal{J}^2\frac{2^{\tilde{q}-1}}{\tilde{q}^2}G^{\tilde{q}}\right)~,\\
    \tilde{I}_{\text{UV}}&=&\frac{1}{2}\int_{0}^{\beta}\int_{0}^{\beta} d\tau_1 d\tau_2\left( \delta(\tau_{1} - \tau_2)\partial_{\tau_2} G - \mathcal{J}^2\frac{2^{q-1}}{q^2}G^{q}\right)~.\label{UV action}
\end{eqnarray}
The CFT action (\ref{CFT action deep IR}) is the same as the action \eqref{CFT action} discussed in section \ref{sec: Brief review of the SYK model} upon  making the replacements $\mathcal{J} \to s\mathcal{J}$ and $q \to \tilde{q}$. The UV action, $\tilde{I}_{\text{UV}}$, has an additional term as compared to that of the undeformed SYK model. Note that so far all we have done is to rewrite (\ref{deformed bilocal action}). We rewrite it in this way because we would like to view $\tilde{I}_{\text{UV}}$ as a perturbation to $\tilde{I}_{\text{CFT}}$ and will be interested in computing its leading effect. 

We have a continuous family of saddle solutions of $\tilde{I}_{\text{CFT}}$ written in terms of reparameterisations, $\phi(\tau)$,  of the circle to itself with a single unit of winding
\begin{equation}
\label{full two point on circle deep IR}
    G_{\phi}(\tau_1,\tau_2) = \phi'(\tau_1)^{\Delta}\,b \, \mbox{sgn}(\tau_1-\tau_2)\left(\frac{\pi}{\beta s \mathcal{J} \sin\left(\frac{\pi(\phi(\tau_1)-\phi(\tau_2))}{\beta}\right)}\right)^{2\Delta} \, \phi'(\tau_2)^{\Delta}~,
\quad\quad \Delta \equiv 1/\tilde{q}~, 
\end{equation}
where the constant $b$ is given by \eqref{constant b}.

We now argue that the leading correction to $\tilde{I}_{\text{CFT}}$ due to the effect of $\tilde{I}_{\text{UV}}$ takes the form of a Schwarzian action and gives a linear-in-temperature contribution to the specific heat. The argument we make is analogous to the one used for the single SYK \cite{Kitaev:2017awl, Rosenhaus:2018dtp}.\footnote{In Appendix \ref{app:Schwarzian contribution to the specific heat} we show that this argument gives the correct low temperature entropy in the integrable case of a single SYK model with $q=2$.} For an alternative treatment of the Schwarzian action and near conformal perturbations see \cite{Jevicki:2016ito, Das:2020kmt}.
It will be convenient to rewrite the reparameterisation modes $\phi(\tau)$ in terms of modes on the line $f(\tau)$, defined by
\begin{equation}
    f(\tau) = \tan\left(\frac{\pi \phi(\tau)}{\beta}\right)~.
\end{equation}
After this transformation we find our solutions \eqref{full two point on circle deep IR} are parameterised as
\begin{equation}
\label{full two point on line}
    G_{f}(\tau_1,\tau_2) = \frac{b}{(s\mathcal{J})^{2\Delta}}\frac{f'(\tau_1)^{\Delta}f'(\tau_2)^{\Delta}}{|f(\tau_1)-f(\tau_2)|^{2\Delta}}~.
\end{equation}
We will want to use \eqref{full two point on line} in $\tilde{I}_{\text{UV}}$, so that we only pick out contributions to the path integral along the conformal saddle solutions. We expand $G_{f}(\tau_1,\tau_2)$ around $(\tau_{1},\tau_{2})=(\tau_{+},\tau_{+})$, where $\tau_{+} \equiv (\tau_1+\tau_2)/2$, giving a series in powers of $\tau_{12} \equiv \tau_1 - \tau_2$,
\begin{equation}
\label{Scwharzian expansion}
    G_{f}(\tau_1,\tau_2) = \frac{1}{(s\mathcal{J})^{2\Delta}|\tau_{12}|^{2\Delta}}\left(1+\frac{\Delta}{6}\tau_{12}^{2} \, \textrm{Sch}(f(\tau_{+}),\tau_{+}) + \mathcal{O}
(\tau_{12}^3)\right)~,
\end{equation}
where the Scwharzian derivative is defined by
\begin{equation}
    \textrm{Sch}(f(\tau_{+}),\tau_{+}) \equiv \frac{f'''(\tau_{+})}{f'(\tau_{+})} - \frac{3}{2}\left(\frac{f''(\tau_{+})}{f'(\tau_+)}\right)^{2} 
    = \frac{1}{2}\left(\left(\frac{2\pi}{\beta}\right)^{2}\phi'(\tau_+)^{2} - \left(\frac{\phi''(\tau_+)}{\phi'(\tau_+)}\right)^{2}\right)~.
\end{equation}
We now substitute the expansion \eqref{Scwharzian expansion} into $\tilde{I}_{UV}$ while changing the integration variables from $(\tau_1,\tau_2)$ to $(\tau_{+},\tau_{12})$. Due to the periodicity of our fields in $\beta$ we can take the new region of integration as $0\leq\tau_{12}< \beta$ and $0\leq\tau_{+}< \beta$. We then carry out the integral over $\tau_{12}$ by taking a cutoff at short time scales beyond $\tau_{12}=\varepsilon/s\mathcal{J}$, where $\varepsilon$ is a small positive number (the range of integration is taken to be $\varepsilon/{s\mathcal{J}}\leq\tau_{12}< \beta-\varepsilon/{s\mathcal{J}}$). Assuming $n \equiv q/\tilde{q}\neq3/2$, we find a term proportional to the Schwarzian action in terms of the cutoff $\varepsilon$
\begin{equation}
\label{UV saddle general n}
    \tilde{I}_{\text{Sch}} = \left[\left(\frac{bn (n-q)\varepsilon ^{1-2\Delta}}{6 q^2}\right)\frac{1}{s\mathcal{J}} -\left(\frac{n}{2n-3}\frac{(2b)^q \varepsilon ^{3-2 n}}{24q^2 s^2}\right)\frac{1}{s\mathcal{J}}\right] \int_{0}^{\beta}\;d\tau_{+}\; \mathrm{Sch}(f(\tau_+),\tau_+)~.
\end{equation}
Here, we have kept only terms in the coefficient of the Schwarzian that are constant in $\beta$ as these contribute to the linear-in-temperature specific heat when the Schwarzian is evaluated on shell. 

The first term in the Schwarzian coefficient (\ref{UV saddle general n}) stems from the kinetic term in $I_{\text{UV}}$, while the second from the non-kinetic term in $I_{\text{UV}}$. Notice that in the large $q$ limit the cutoff dependence of the first term goes like $\varepsilon$ whilst that of the second term goes like $\varepsilon^{3-2n}$. This suggests that for $n$ close to 1 both terms are important as we take the cutoff $\varepsilon \to 0$. For larger values of $n$, the second term dominates.\footnote{Since the coefficient of the Schwarzian governs the linear-in-temperature specific heat, this competition of factors could perhaps underlie the transition we see in the value of $\bar{\aleph}(s,n)$ for small values of $n$ in figure \ref{fig:aleph_bar}.} For the sake of concreteness, let us focus on the case $n=2$. Equation \eqref{UV saddle general n} becomes
\begin{equation}
\label{UV saddle}
   \tilde{I}_{\text{Sch}} = \left[\left(\frac{b (2-q)\varepsilon ^{1-2\Delta}}{3 q^2}\right)\frac{1}{s\mathcal{J}} -\left(\frac{(2b)^q}{12 q^2 s^2  \varepsilon }\right)\frac{1}{s\mathcal{J}}\right] \int_{0}^{\beta}\;d\tau_{+}\; \mathrm{Sch}(f(\tau_+),\tau_+)~.
\end{equation}
The non-kinetic term goes like $1/\varepsilon$ and so provides the most significant correction to the conformal part of the action. The reparametrisation symmetry is broken by choosing the saddle of the Schwarzian which occurs when $\phi(\tau)=\tau$. Substituting this into \eqref{UV saddle} we find the linear-in-temperature contribution to the entropy to leading order in $\varepsilon$
\begin{equation}
    \frac{S_{\text{Sch}}}{N} = \frac{(2b)^q}{6 q^2 s^2 \varepsilon }\left(\frac{2\pi^{2}}{s \beta \mathcal{J}}\right)~.
\end{equation}

The takeaway message of this analysis is that due to the dominance of the second term \eqref{UV saddle} the correction to the conformal action comes from the strongly coupled phase of the theory rather than the weakly coupled UV regime which is customary for the undeformed SYK model. Holographically, for those deformed SYK models having both an intermediate and deep IR near-fixed point, we anticipate the emergence of the Schwarzian mode in the interior of an asymptotically AdS$_2$ spacetime  flowing to a distinct infrared AdS$_2$ region. 

\subsection{Schwarzian for the intermediate IR}
\label{Schwarzian for the intermediate IR}
We now proceed to consider the conformal fixed point associated to $H_{q}$ with a small perturbation near the fixed point. By taking $\Sigma \to \Sigma + \partial_{\tau}$  in \eqref{deformed bilocal action} we can then write $I = I_{\text{CFT}} + I_{\text{pert}}$ where
\begin{eqnarray} \label{CFT action intermediate IR}
    I_{\text{CFT}}&=&-\frac{1}{2}\log\det(-\Sigma) + \frac{1}{2}\int_{0}^{\beta}\int_{0}^{\beta}d\tau_1 d\tau_2 \left(\Sigma G - \mathcal{J}^2\frac{2^{q-1}}{q^2}G^{q} \right)~,\\
    I_{\text{pert}}&=&\frac{1}{2}\int_{0}^{\beta}\int_{0}^{\beta} d\tau_1 d\tau_2 \left( \delta(\tau_{1}-\tau_2)\partial_{\tau_2} G - s^2\mathcal{J}^2\frac{2^{\tilde{q}-1}}{\tilde{q}^2}G^{\tilde{q}} \right)~.\label{pert action intermediate IR}
\end{eqnarray}
As in the previous section, we make an expansion of the saddle solution to $I_{\text{CFT}}$ in powers of $\tau_{12}$, written in terms of soft modes $f(\tau_+)$,
\begin{equation}
\label{Scwharzian expansion intermediate}
    G_{f}(\tau_1,\tau_2) = \frac{1}{\mathcal{J}^{2\Delta}|\tau_{12}|^{2\Delta}}\left(1+\frac{\Delta}{6}\tau_{12}^{2} \, \textrm{Sch}(f(\tau_{+}),\tau_{+}) + \mathcal{O}(\tau_{12}^3)\right)~,
\end{equation}
where now $\Delta = 1/q$. Substituting this into $I_{\text{pert.}}$, we change variables to $(\tau_{12},\tau_{+})$ and carry out the $\tau_{12}$ integral with a short time scale cutoff $\varepsilon/\mathcal{J}$. Keeping only terms constant in $\beta$ (since these are the terms that contribute to the linear-in-temperature part of the entropy when the Schwarzian is evaluated on shell). Again focusing on $n=2$ for the sake of concreteness, we find  
\begin{equation}
    I_{\text{Sch }} = \left[\left(\frac{b (1-q)\varepsilon ^{1-2\Delta}}{6 q^2}\right)\frac{1}{\mathcal{J}} \right]\int_{0}^{\beta}\;d\tau_{+} \;\textrm{Sch}(f(\tau_+),\tau_+)~.
\end{equation}
The coefficient of the Schwarzian is seen to come purely from the kinetic term in (\ref{pert action intermediate IR}), mimicking the behaviour of the underformed SYK model with Hamiltonian $H_{q}$. Accordingly, the linear-in-temperature term in the entropy expansions \eqref{Intermediate IR entropy} and \eqref{entropy large q} are found to be the same, and do not depend on $s$.

\subsection{Conformal relevant perturbation theory}
\label{CPT}
We would now like to test the hypothesis that the leading infrared correction away from conformality of the intermediate IR phase can be studied using conformal perturbation theory. The starting point  \cite{Tikhanovskaya:2020elb,Cruz:2022uic} is to view the deformed SYK model near the intermediate fixed point as a conformal field theory perturbed by a series of relevant primary operators $O_{h}(\tau)$ of weight $h \in (0,1)$.\footnote{Since in one dimension we can conformally map the line to the circle, we can employ conformal perturbation theory methods on the circle.} More explicitly,
\begin{equation}
\label{CPT action}
    I =I_{\text{CFT}}+\sum_{h \in \text{rel.}} g_{h} \int_{0}^{\beta} d \tau \, O_{h}(\tau)~,
\end{equation}
where $h$ denotes the conformal weight of the given operator. We note here that the spectrum of conformal operators discussed in \cite{Polchinski:2016xgd, Maldacena:2016hyu, Gross:2016kjj, Tikhanovskaya:2020elb,Cruz:2022uic}, does not contain any relevant operators with $h \in (0,1)$. They are in fact all irrelevant and are encoded in the operator product expansion of the fusion of two fermionic operators. Motivated by the structure of the Hamiltonian deformation (\ref{deformed Hamiltonian}), here we will focus instead on the following microscopic operator
\begin{equation}
\label{relevant operator def}
O_{h}(\tau) \equiv \mathcal{N}_h  \, { i^{\frac{\tilde{q}}{2}} } \sum_{1\leq i_1 <\cdot\cdot\cdot< i_{\tilde{q}}\leq N}    J_{i_1i_2\cdot\cdot\cdot i_{\tilde{q}}}\psi_{i_1}\psi_{i_2}\cdot\cdot\cdot \psi_{i_{\tilde{q}}}~. 
\end{equation}
This operator is to be understood in an averaged sense since it depends on the couplings $J_{i_1i_2\cdot\cdot\cdot i_{\tilde{q}}}$ which are averaged over.\footnote{This is somewhat in the spirit of \cite{Belin:2020hea}. It is interesting that in contrast to operators associated to large black holes, which are highly irrelevant, ${O_{h}(\tau)}$ is a complicated operator that is relevant. Perhaps, given its averaged nature, one can associate an entropy different from that of the horizon to its effect on the bulk spacetime.} The operator ${O_{h}(\tau)}$ involves a product of $\tilde{q}$ fermions. In the undeformed model each fermion has scaling dimension $\Delta_\psi =1/q$, so the naive estimate of the total weight of the operator (\ref{relevant operator def}) is $h=1/n$  up to small corrections, which is within the relevant window $h\in(0,1)$. We fix the value of $\mathcal{N}_h$ implicitly by our choice of normalisation for the conformal two-point function averaged over the couplings
\begin{equation}\label{conformal two-point function}
{\left\langle O_{h}\left(\tau_{1}\right) O_{h}\left(\tau_{2}\right)\right\rangle_{\beta}} = N \left(\frac{\pi }{\beta  \mathcal{J} \sin \left(\frac{\pi  \tau_{12}}{\beta }\right)}\right)^{2 h}~.
\end{equation}
The action $I_{\text{CFT}}$ in (\ref{CPT action}) governs the intermediate IR fixed point. According to conformal perturbation theory we find the following free energy  
\begin{equation}
\label{CPT free energy}
    {\beta F} = {\beta F_{\text{CFT}}} +g_{h} \int_{0}^{\beta} d \tau {\left\langle O_{h}\right\rangle_{\beta}}-\frac{g^2_{h} }{2}\int_{0}^{\beta} \int_{0}^{\beta} d \tau_{1} d \tau_{2}{\left\langle O_{h}\left(\tau_{1}\right) O_{h}\left(\tau_{2}\right)\right\rangle_{\beta}} + \cdots~.
\end{equation}
Here $O_h$ is the relevant operator (\ref{relevant operator def}), and again it is understood that we are averaging over the couplings. The one-point function of the $O_h$ vanishes under the assumption of conformal invariance of the vacuum. Using the conformal form of the two-point function \eqref{conformal two-point function}, the second order correction is given by \cite{Tikhanovskaya:2020elb,Cruz:2022uic,Delacretaz:2021ufg} 
\begin{equation}
\label{CPT free energy correction}
    -\frac{\beta \delta^{2} F_{h}}{N}=\frac{\pi^{2 h-\frac{1}{2}} \Gamma\left(\frac{1}{2}-h\right)}{2 \Gamma(1-h)} \frac{g_{h}^{2} }{ \mathcal{J}^{2}(\beta \mathcal{J})^{2 h-2}}~.
\end{equation}
We will now provide evidence that the above correction indeed gives the leading correction to the intermediate CFT as we flow towards the IR. First, we consider the large $q$ limit with $n=2$, where we have the analytical form of the correction. We then consider general $n$ in the large $q$ limit and at finite $q$, where we compare to numerics. 
\newline\newline
\textbf{Case I: $n=2$.} The intermediate IR CFT free energy is known analytically \cite{Anninos:2020cwo} at large $q$ with ${q}/\tilde{q} = 2$. Concretely, in the regime $1 \ll \beta\mathcal{J} \ll 1/s^2$, the free energy of the deformed model can be written as
\begin{equation}
\label{free energy split}
    -\frac{\beta F}{N} =  \left[\frac{1}{q^{2}}\beta \mathcal{J} +\left(S^{\mathrm{free}}_0 -\frac{\pi^2}{4 q^2} \right) + \frac{\pi^{2}}{2 q^2}\frac{1}{\beta \mathcal{J}} + \cdots\right] + \left[ \frac{2s^{2}}{q^2}\beta\mathcal{J}\log \left( \frac{2\beta\mathcal{J}}{\pi} \right)+ \cdots\right]~,
\end{equation}
where the terms in the first square bracket are derivable from $I_{\text{CFT}}$ given by (\ref{CFT action intermediate IR})  accompanied by the leading irrelevant operators \cite{Tikhanovskaya:2020elb,Cruz:2022uic}, and they grow with increasing temperature. The terms in the second square bracket stem from the corrections due to relevant operators. 

\begin{figure}[H]
        \centering
         \subfigure[$s^2=10^{-6}$]{
                \includegraphics[height=4.55cm]{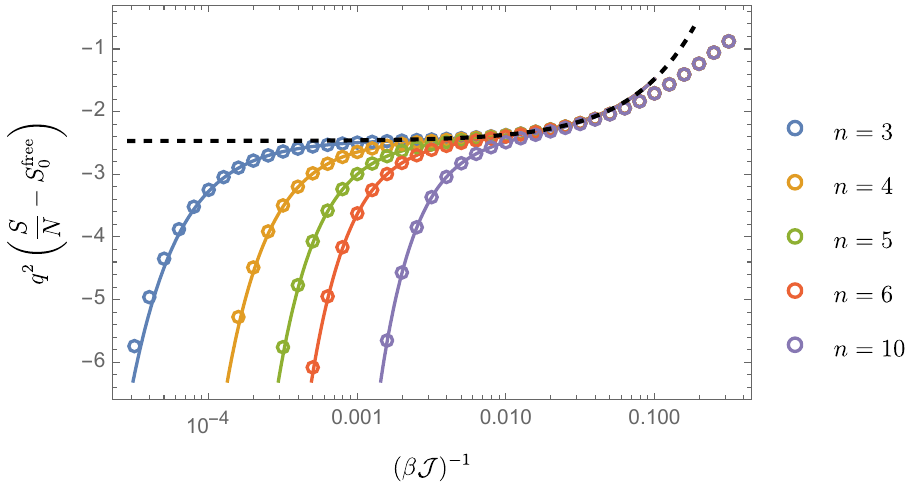}\label{fig:s2_-6}}  \quad
        \subfigure[$s^2=10^{-8}$]{
                \includegraphics[height=4.55cm]{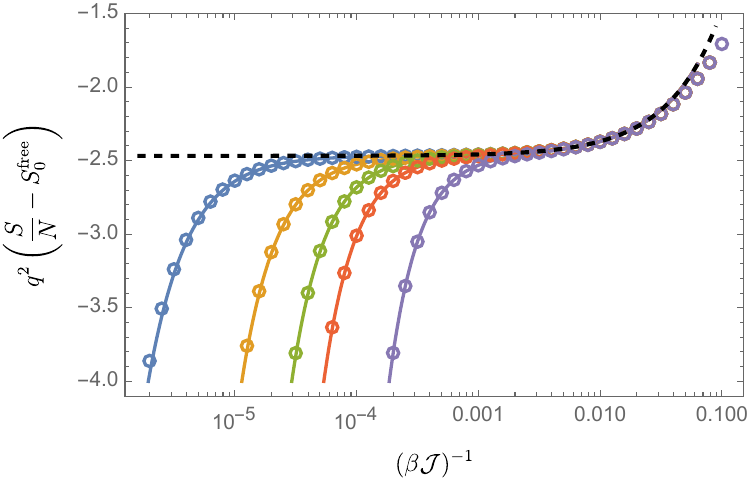} \label{fig:s2_8}} 
                 \caption{Entropy as a function of temperature (in logarithmic scale) for the intermediate IR phase in the large $N$ and $q$ expansion. The circles give numerical results. The solid lines give the analytical prediction \eqref{Intermediate IR general n} with both the leading irrelevant and relevant corrections. The dashed line gives the analytical prediction \eqref{Intermediate IR general n} with only the leading irrelevant correction.}
\label{fig:intermediate_IR}
\end{figure}

We will now argue that the leading relevant correction to the free energy arises from a relevant operator of weight $h=1/2$. Given that expression \eqref{CPT free energy correction} diverges when we take $h =1/2$,  we are led to a divergent contribution to the free energy which requires regularisation. As a simple regularisation scheme, we take $h =1/2 - h_\varepsilon$ for some small number $h_\varepsilon>0$, such that 
\begin{equation}
    -\frac{\beta \delta^{2} F_{h=1/2-h_\varepsilon}}{N}=\frac{\Gamma(h_\varepsilon)^{2} (g^2_{1/2}/\mathcal{J}^2)}{4\Gamma(2h_\varepsilon)}\beta\mathcal{J}\left(\frac{2\beta\mathcal{J}}{\pi}\right)^{2h_\varepsilon}~.
\end{equation}
Expanding in small $h_\varepsilon$ gives
\begin{equation}
-\frac{\beta \delta^{2} F_{h=1/2}}{N} = \frac{g^2_{1/2}}{2h_\varepsilon\mathcal{J}^2}\beta \mathcal{J} + \frac{g^2_{1/2}}{\mathcal{J}^2}\beta \mathcal{J}\log \left(\frac{2 \beta\mathcal{J}}{\pi}\right) + \mathcal{O}(h_\varepsilon)~.
\end{equation}
Consequently, the divergent term only affects the zero point energy whose contribution to the free energy is independent of $\beta$. 
The remaining $h_\varepsilon$-independent terms agree with \eqref{free energy split} provided we take
\begin{equation}
g^{2}_{1/2} \to \frac{2 s^2\mathcal{J}^2}{q^2} \quad \mbox{as}\quad q\to\infty~.
\end{equation}
This provides evidence that for $n = 2$, and in the large $q$ limit, we can view $O_h$ in (\ref{relevant operator def}) as a relevant conformal primary of conformal dimension $h= \tilde{q}/q = 1/2$. We now consider the case of general $n$. 
$\,$\newline\newline
\textbf{Case II: General $n$.} For general $n$ we do not have access to an analytic form of the free energy near the intermediate IR fixed point. Nonetheless, we can test the prediction from conformal perturbation theory against numerical results. To do so, we compute the entropy of the model numerically in the large $q$ limit with $q=n\tilde{q}$ as described in section \ref{Large q limit of the deformed SYK with general n}. Taking $h =1/n$ in \eqref{CPT free energy correction} and using the formula $S=(1-\beta\partial_{\beta})(-\beta F)$ we find that the correction to the entropy due to the relevant perturbation is given by
\begin{equation} \label{CPT correction}
\frac{\delta^{2} S_{h=1/n}}{N} = \left(1-\left(2-\frac{2}{n}\right)\right) \frac{\pi ^{\frac{2}{n}-\frac{1}{2}} \Gamma \left(\frac{1}{2}-\frac{1}{n}\right)(g_{1/n}^{2}/\mathcal{J}^2)}{2 \Gamma \left(1-\frac{1}{n}\right)}\left(\beta\mathcal{J}\right)^{2-\frac{2}{n}}~.
\end{equation}
From this it follows that the entropy near the intermediate IR fixed point, as predicted by conformal perturbation theory, can be expressed as
\begin{equation}
\label{Intermediate IR general n}
   q^{2}\left(\frac{S}{N}-S^{\mathrm{free}}_{0}\right)=\left[-\frac{\pi^{2}}{4} +\frac{\pi^{2}}{\beta \mathcal{J}} + \cdots \right] + \left[q^{2}\frac{\delta^{2} S_{h=1/n}}{N} + \cdots \right]~.
\end{equation}
The terms in the first square bracket are derivable from $I_{\text{CFT}}$ and the irrelevant operators whilst the terms in the second square bracket are proposed to come from the relevant deformation. In figure \ref{fig:intermediate_IR} we plot numerical results for the entropy in the intermediate IR phase against the analytic prediction \eqref{Intermediate IR general n}, as well as the linear-in-temperature curve without the correction from the relevant perturbation. We show plots for $s^2=10^{-6}$ and  $s^2=10^{-8}$, both with curves for $n=3,4,5,6$ and $10$. Provided
\begin{equation} \label{large q CPT coupling}
g^{2}_{1/n} \to \frac{n^{2} s^2\mathcal{J}^2}{2q^2} \quad \mbox{as}\quad q\to\infty~,
\end{equation}
there is strong agreement with the numerics. 

We can also study higher order corrections from conformal perturbation theory. By dimensional analysis the $k^{\text{th}}$ order correction is found to be of the form
\begin{equation}
\frac{\delta^{k}S_{h=1/n}}{N} \propto s^{k}(\beta\mathcal{J})^{k-\frac{k}{n}},\quad\quad k\geq2~.
\end{equation}
To find the sub-leading relevant correction we subtract the prediction \eqref{Intermediate IR general n}, up to and including the leading relevant correction, from the numerically calculated entropy and perform a numerical fit. For the values of $n$ we have tested we find the sub-leading relevant correction to be  proportional to $s^{4}(\beta\mathcal{J})^{4-\frac{4}{n}}$. We also find evidence, as discussed below, that this is true even at finite $q$. The absence of a term proportional to  $s^{3}(\beta\mathcal{J})^{3-\frac{3}{n}}$ leads us to believe that that the conformal three-point function is sub-leading in the large $N$ expansion, as is seen to be the case for the conformal three-point functions discussed in \cite{Gross:2017aos}. 

\begin{figure}[H]
    \centering
    \vspace{0mm}
    \includegraphics[width=0.6\columnwidth]{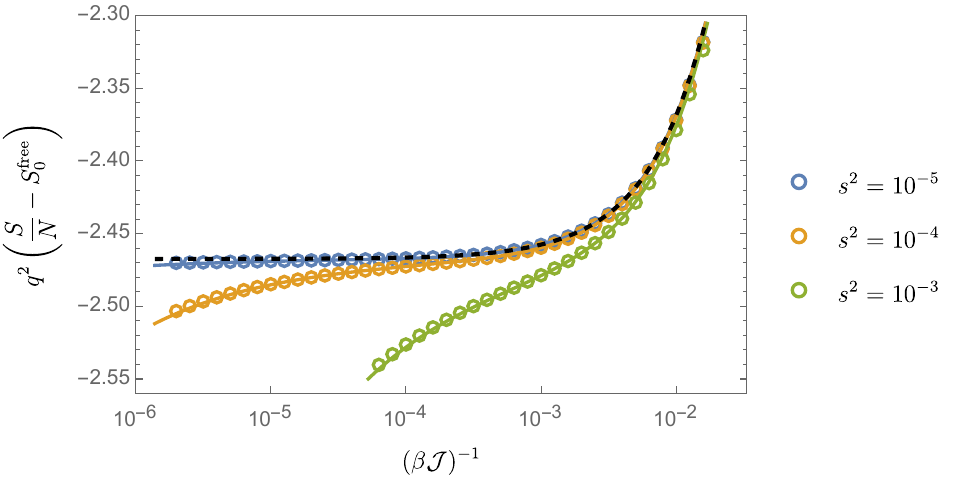}
    \caption{Entropy as a function of temperature (in logarithmic scale) for $n=1.3$ and $s^2=10^{-5}, 10^{-4}, 10^{-3}$ in the intermediate IR phase in the large $N$ and $q$ expansion. The circles give numerical results. The solid lines give the analytical prediction \eqref{Intermediate IR general n} with both the leading irrelevant and relevant corrections. The dashed line gives the analytical prediction \eqref{Intermediate IR general n} with only the leading irrelevant correction.}\label{fig:n_1_3}
\end{figure}

Finally, it is also interesting to note that we also find an intermediate IR regime for values of $n$ such that $1<n<2$, whose behaviour is in agreement with \eqref{Intermediate IR general n}. In figure \ref{fig:n_1_3} we plot the intermediate IR regime for $n=1.3$ and various values of $s^{2}$, again seeing excellent agreement with the prediction from conformal perturbation theory. From our analysis in section \ref{Large q limit of the deformed SYK with general n} we would also expect the zero temperature entropy of such flows to have a non-trivial $s$ dependence, giving them an additional richness compared to the case $n\geq2$.



\begin{figure}[H]
        \centering
         \subfigure[$q=4, \tilde{q}=2$]{
                \includegraphics[scale=0.5]{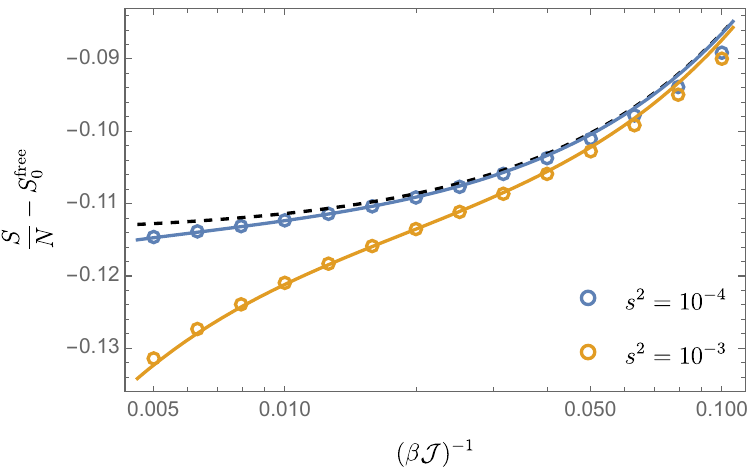}\label{q_2_4_CPT}}  \quad\quad
        \subfigure[$q=8, \tilde{q}=4$]{
                \includegraphics[scale=0.5]{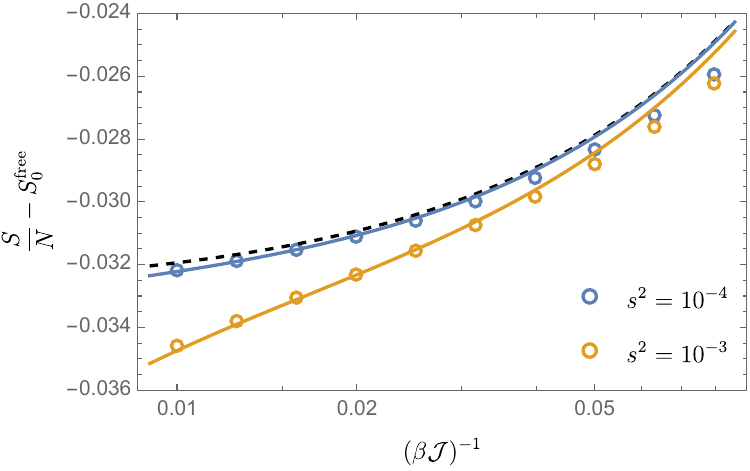} \label{q_4_8_CPT}}  \quad\quad
       \subfigure[$q=6, \tilde{q}=2$]{
                \includegraphics[scale=0.5]{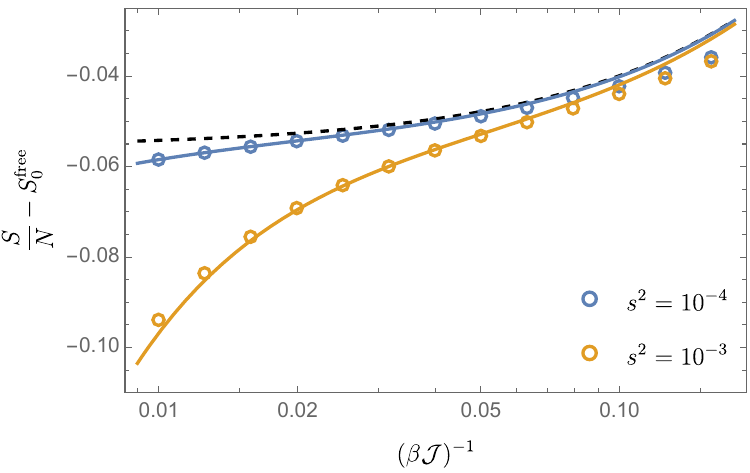} \label{q_2_6_CPT}}  \quad\quad
       \subfigure[$q=8, \tilde{q}=2$]{
                \includegraphics[scale=0.5]{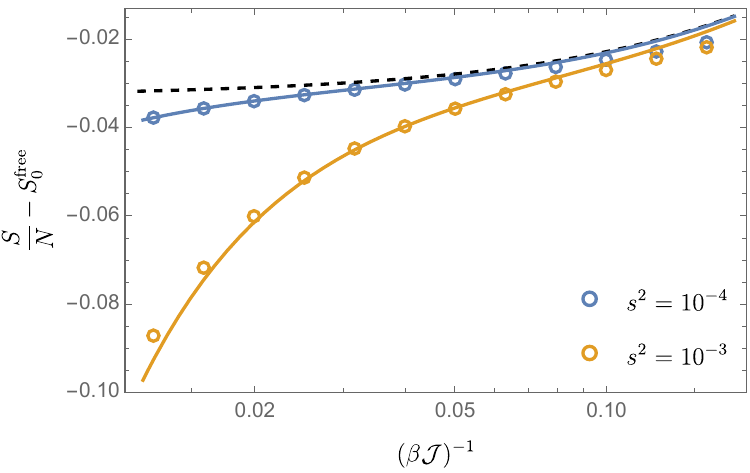} \label{q_2_8_CPT}}                         
                \caption{Entropy as a function of temperature (in logarithmic scale) for the intermediate IR phase at finite $q$ in the large $N$ expansion. The circles give numerical results. The solid lines give the analytical prediction \eqref{finite q CPT prediction} and \eqref{finite q CPT coupling} with values for $\gamma(q, \tilde{q})$ shown in Table \ref{gamma table}. The dashed line gives the analytical prediction \eqref{finite q CPT prediction} with only the leading irrelevant correction.}\label{fig:finite_q_CPT}
\end{figure}

$\,$\newline\newline
\textbf{Case III: Finite $q$.} We now test whether the perturbative correction \eqref{CPT correction} still applies at finite $q, \tilde{q}$ and large $N$. In this case, the predicted entropy near the intermediate IR fixed point is given by
\begin{equation}\label{finite q CPT prediction}
   \frac{S}{N}-S^{\mathrm{free}}_{0}=\left[- \int_{0}^{1/q} d x\; \pi\left(\frac{1}{2}-x\right) \tan \pi x +  \frac{4\pi^{2}\alpha(q)}{\beta \mathcal{J}} + \cdots~ \right] + \left[\frac{\delta^{2} S_{h=1/n}}{N} + \cdots \right]~,
\end{equation}
and the the coupling constant of our conformal operator takes the form 
\begin{equation} \label{finite q CPT coupling}
    g^{2}_{1/n} = \gamma(q, \tilde{q}) s^2\mathcal{J}^2\,,
\end{equation}
where $\gamma(q, \tilde{q})$ is an unknown function which, from  \eqref{large q CPT coupling}, we know tends to $1/(2 \tilde{q}^2)$ in the large $\tilde{q}$ limit. The value of $\gamma(q, \tilde{q})$ can be found by fitting the prediction \eqref{finite q CPT prediction} to numerically determined values for the entropy in the intermediate IR phase. In figure \ref{fig:finite_q_CPT} we plot numerical results against the prediction \eqref{finite q CPT prediction} and \eqref{finite q CPT coupling} with values for $\gamma(q, \tilde{q})$ shown in Table \ref{gamma table}. We show plots with $s^2=10^{-4}$ and $s^2=10^{-3}$.

\begin{table}[H]
\centering
\begin{tabular}{ |c|c|c| } 
 \hline
 $q$ & $\tilde{q}$ & $\gamma(q,\tilde{q})$ \\ 
 \hline
 4 & 2 & 0.098 \\ 
 6 & 2 & 0.111 \\ 
 8 & 2 & 0.116 \\ 
 8 & 4 & 0.028 \\ 
 \hline
\end{tabular}
\caption{Numerical values for $\gamma(q,\tilde{q})$ in \eqref{finite q CPT coupling} found by fitting the prediction \eqref{finite q CPT prediction} to numerically determined values for the entropy in the intermediate IR phase.}\label{gamma table}
\end{table}

As for the large $q$ limit, we can also find the sub-leading relevant correction by performing a numerical fit. In all cases considered, we again find that the sub-leading relevant correction is proportional to $s^4(\beta\mathcal{J})^{4-\frac{4}{n}}$.

\section{Outlook -- Geometrisation of an RG flow}
\label{sec_outlook}
The goal of this paper has been to explore RG flows at strong coupling, and in particular at finite temperature, for deformations of SYK models. We have identified a class of models permitting a robust treatment by means of both numerical and analytic methods. Given the holographic character of SYK models, our analysis opens up an interesting chapter in the story of holographic renormalisation \cite{deBoer:1999tgo,deHaro:2000vlm}, which has so far been  explored mostly from the bulk perspective. We have identified models exhibiting RG flows between two near-fixed points and provided an interpretation in terms of conformal perturbation theory. The general character of the models is a sum of two ordinary SYK Hamiltonians (\ref{deformed Hamiltonian}), but with differing numbers of interacting fermions. As for the ordinary SYK model, the flows we study preserve a rich thermodynamic structure and exhibit an extensive entropy all the way into the deep infrared/small temperature regime. 

Our analysis is performed entirely from the perspective of the microphysical theory. From a holographic perspective, it is interesting to assess what features the putative holographic dual will exhibit. In the vicinity of each near-fixed point, it is natural to postulate that the bulk theory will mimic that of an ordinary SYK, whose thermodynamic features in the large $N$ limit are captured by a JT gravity theory governed by the classical Euclidean action
\begin{equation}
S_E = - \frac{1}{2\kappa} \int_{\mathcal{M}} d^2 x \sqrt{g} \left( \phi R + U(\phi)  \right) -\frac{1}{\kappa} \int_{\partial\mathcal{M}} \sqrt{h} \phi K
\end{equation}
with dilaton potential $U(\phi) = -2\alpha\phi$ with $\alpha$ real and positive. For $U(\phi) = -2\alpha\phi$, one finds that the two-dimensional metric $g_{ij}$ is Euclidean AdS$_2$ at the classical level.  At finite temperature, $\mathcal{M}$ is taken to have a disk topology with $S^1$ boundary $\partial\mathcal{M}$, and we have the metric on the Poincar\'e disk. The thermodynamic properties of asymptotically AdS$_2$ geometries follow readily from the form of $U(\phi)$. The specific heat $C_U$ and temperature, for instance, are given by \cite{Grumiller:2007ju, Anninos:2017hhn,Witten:2020ert} 
\begin{equation}
C_U =  \frac{2\pi}{\kappa} \frac{U(\phi_h)}{\partial_\phi U(\phi_h)}~, \quad\quad \beta = \frac{4\pi}{U(\phi_h)}~,
\end{equation}
where $\phi_h$ is the value of the dilaton field $\phi$ at the Euclidean horizon. It follows that the near-fixed point exhibits a specific heat linear in the temperature. 

For two near-fixed points, the ratio of the specific heats fixes the ratio, $\mathcal{R} \equiv \alpha_{\text{UV}} / \alpha_{\text{IR}}$, of the slopes for the two linear regimes of $U(\phi)$. For the models we have studied, we find $\mathcal{R} > 1$. This is in line with an increasing number of degrees of freedom as we go to higher temperatures and can be viewed as a consequence of unitarity and thermal equilibrium. Interestingly, as we increase the temperature, we remove an increasingly large portion of the interior AdS$_2$. At large enough temperatures, the remaining geometry becomes a pure AdS$_2$ and there is boundary soft mode governed by the Schwarzian action. This is the bulk dual of the Schwarzian associated to the intermediate near-fixed point discussed in section \ref{Schwarzian for the intermediate IR}. 

\begin{figure}[H]
        \centering
         \subfigure[$\beta \mathcal{J} \gtrsim (\beta \mathcal{J})_*$]{
                \includegraphics[scale=0.5]{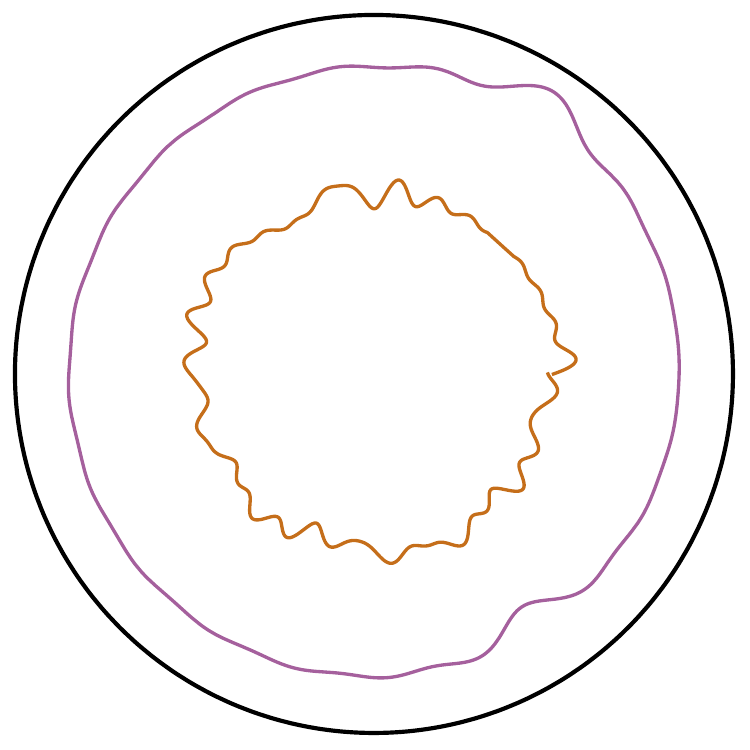}\label{ads2sch}}  \quad\quad
       \subfigure[$\beta \mathcal{J} \lesssim (\beta \mathcal{J})_*$]{
                \includegraphics[scale=0.5]{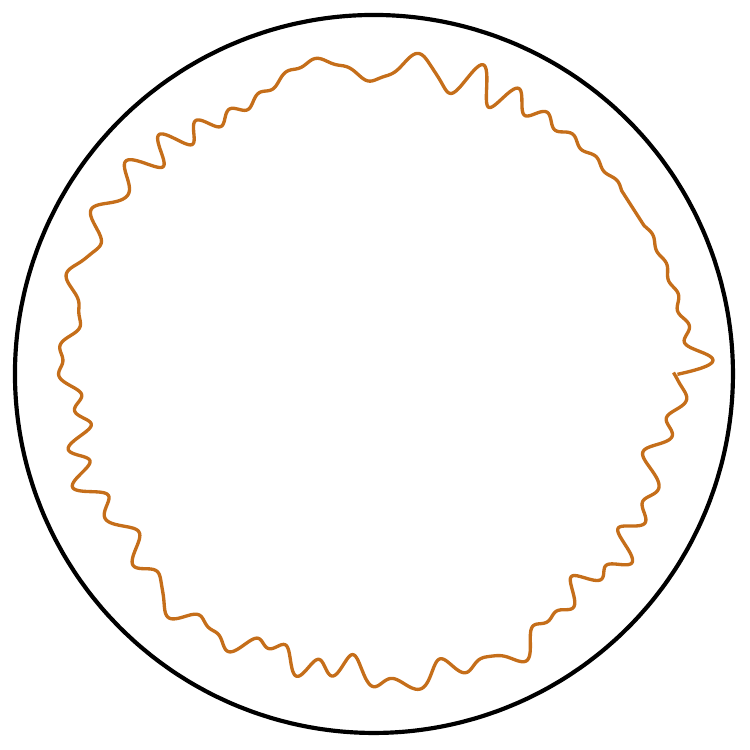} \label{ads2sch2}}                         
                \caption{Pictorial representation of the two Schwarzian soft modes appearing inside Euclidean AdS$_2$. (a) For $\beta \mathcal{J}$ larger that some critical $(\beta \mathcal{J})_*$ there is a Schwarzian soft mode appearing in the deep interior of AdS$_2$. (b) For $1 \ll \beta \mathcal{J} \lesssim (\beta \mathcal{J})_*$, there is a Schwarzian soft mode residing closer to the AdS boundary. In the large $q$ model with $n=2$, $(\beta \mathcal{J})_* \sim s^{-2}$, with $s \ll 1$. }\label{fig:ads2}
\end{figure}

Continuity of the thermodynamic quantities along the RG flow, throughout which the theory remains in the strongly coupled phase, suggests that the geometric picture continues to hold between the two near-fixed points. For this to occur, one can  invoke  \cite{Anninos:2020cwo} a more general dilaton potential $U(\phi)$, as studied for example in \cite{Grumiller:2007ju, Anninos:2017hhn,Witten:2020ert,Maxfield:2020ale} with linear behaviour at the two endpoints. The classical geometry will be asymptotically, but not isometrically, Euclidean AdS$_2$. The presence of a macroscopic entropy in the deep infrared/low temperature regime of the flow leads us to  postulate that the dual geometry retains a horizon. In section \ref{sec_int_IR} we  argued that the RG flow is triggered by a relevant operator of weight $\Delta_{\text{rel}} = \tilde{q}/q < 1$. Thus, the bulk theory should have a corresponding field associated to the relevant operator. Moreover, as one flows to the interior of the geometry an additional AdS$_2$ region emerges, corresponding to the near-fixed point in the deep infrared. Associated to this is the presence of a soft mode residing at the boundary of the near-AdS$_2$ geometry in the deep finite interior region, governed by the Schwarzian action. It is interesting that this soft-mode resides entirely within the geometric description.\footnote{A similar emergence of a soft-mode in the interior of an interpolating geometry was also discussed for the centaur geometries studied in \cite{Anninos:2018svg,Anninos:2022hqo,Anninos:2022ujl}.} We depict this phenomenon in figure \ref{fig:ads2}. The appearance of a worldline theory in the midst of a gravitating spacetime is a phenomenon worth pursuing in more detail.  


Looking forward, it will be interesting to test the hypothesis that the microscopic RG flow is captured by a  dilaton-gravity theory with generalised dilaton potential by computing other observables such as the correlation functions of the fermionic operators. Moreover, one can consider larger classes of deformations. A particular family of such deformations  is given by  concatenating multiple SYK Hamiltonians
\begin{equation}
{H}_{\text{tot}} = \sum_{i=1}^k  \lambda_i {H}_{q_i}~,
\end{equation} 
with $q_1 > q_2 > \ldots > q_k$ and $\lambda_i \in \mathbb{C}$. Although unitarity enforces $\lambda_i \in \mathbb{R}$, it is interesting to consider the more general complex case, as such models make contact with the physics of open quantum systems \cite{Liu:2020fbd,Garcia-Garcia:2021rle,Bentsen:2021ukm} which, in turn, may bear relevance to the problem of de Sitter. The case $k=3$ is particularly interesting, given the recent realisation \cite{Anninos:2022hqo} of a thermodynamically stable macroscopic portion of dS$_2$ suspended between two approximately AdS$_2$ geometries, one near the boundary and the other in the deep interior. Technically, this  requires reaching lower temperatures in the numerical computations. This might be achieved by incorporating new techniques such as spectral \cite{Cruz:2022uic} or Krylov \cite{Kobrin:2020xms} methods and/or new approximate models such as the sparse models studied in \cite{Xu:2020shn, Garcia-Garcia:2020cdo, Tezuka:2022mrr}. Building a microphysical model\footnote{Rearrangements of a microscopic dual of quantum AdS to obtain dS microstates also plays an interesting role  in the approach of \cite{Shyam:2021ciy,Coleman:2021nor,Silverstein:2022dfj}, and also \cite{Susskind:2021esx,Ecker:2022vkr}.
} for two-dimensional de Sitter from the ingredients of SYK, as originally envisioned in \cite{Anninos:2017hhn,Anninos:2018svg}, is left to near-future work.



\section*{Acknowledgements}

It is a pleasure to acknowledge Alexandre Belin, Nikolay Bobev, Shira Chapman, Luca Delacretaz, Masanori Hanada, Eleanor Harris, Diego Hofman, Beatrix M\"uhlmann, Ben Pethybridge, Andrew Scull, and David Vegh for useful discussions. 
D.A. is funded by the Royal Society under the grant ``The Atoms of a deSitter Universe". The work of D.A.G. is funded by UKRI Stephen Hawking Fellowship ``Quantum Emergence of an Expanding Universe". S.U.S.  is funded by the Royal Society under the grant ``The Resonances of a deSitter Universe". D.A. would like to thank the BEL center, KU Leuven and ULB for their kind hospitality during the completion of this work. D.A.G. would like to further thank the University of Amsterdam, the University of Kentucky and the Perimeter Institute for kind hospitality during the completion of this work. Research at Perimeter Institute is supported by the Government of Canada through the Department of Innovation, Science and Economic Development and by the Province of Ontario through the Ministry of Colleges and Universities. We also acknowledge the use of King's Computational Research, Engineering and Technology Environment (CREATE) \cite{kings}.



\appendix

\section{Numerical computation of $\alpha{(q)}$ in a single SYK model}
\label{app: alpha}

In section \ref{sec: Brief review of the SYK model}, we saw that the entropy of the single SYK model has a small temperature expansion given by
\begin{equation}
\label{finite q entropy appendix}
    \frac{S}{N} = \left(S^{\mathrm{free}}_0 - \int_{0}^{1/q} d x\; \pi\left(\frac{1}{2}-x\right) \tan \pi x\right) +  \frac{4\pi^{2}\alpha(q)}{\beta \mathcal{J}} + \cdots~.
\end{equation}
Here we describe how to compute the coefficient $\alpha(q)$ numerically. The first step is to numerically compute the large $N$ entropy, $S/N$, of the model at a single low temperature point. We then subtract off the temperature independent piece of \eqref{finite q entropy appendix} and multiply the answer by $\beta\mathcal{J}/(4\pi^2)$ to obtain a value for $\alpha(q)$ up to corrections of order $(\beta\mathcal{J})^{-2}$.  To find the entropy, we numerically solve the Schwinger-Dyson equations with $s=0$, as shown in Appendix \ref{appendix: numerical}. In figure \ref{fig:alpha_q} we plot the numerical values of $\alpha(q)$ and compare it with the two-sided Pad\'e approximant, found in \cite{Tarnopolsky:2018env}, 
\begin{equation}
\label{pade}
\alpha(q) = \frac{3 (3 \pi -2) q+\pi ^2-18 \pi +24}{6 q^2 \left(2 (3 \pi -2) q+\pi ^3+8\right)} \,.
\end{equation}
Given the agreement with the numerics, we directly use \eqref{pade} in our numerical computations.
 \begin{figure}[H]
    \centering
    \vspace{0mm}
    \includegraphics[width=0.5\columnwidth]{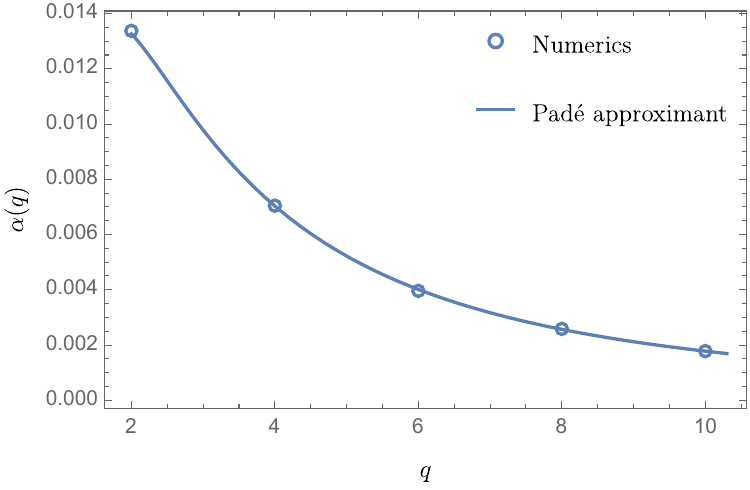}
    \caption{The coefficient $\alpha(q)$ as a function of $q$. The circles are numerical computations while the solid blue curve is given by the Pad\'e approximation \eqref{pade}. We used $\beta\mathcal{J} \sim 10^{2.3}$ for the numerical computations.}\label{fig:alpha_q}
\end{figure}

\section{Small $s$ expansion for $\bar{\aleph}(s,n)$}
\label{section: Towards an analytical form for alephbar}
In this appendix, we provide an analytic form for $\bar{\aleph}(s,n)$, when $n\geq2$ and $s \ll 1$ by fitting the numerical data. For $n=2$, we know $\bar{\aleph}(s,n=2)$ analytically and it is given by $\bar{\aleph}$ in equation (\ref{aleph bar}). It is straightforward to obtain
\begin{equation}
\bar{\aleph}(s,n=2) \to \frac{1/2}{s} + \cdots \,,
\end{equation}
in the small $s$ expansion. Given the shape of the curves from the numerical results, we propose the following structure for general $n$ in the small $s$ limit,
\begin{equation} \label{small s}
\bar{\aleph}(s,n) \to \frac{a(n)}{s^{b(n)}} + \cdots \,,
\end{equation}
where $a(n), b(n)$ can depend on $n$ but are independent of $s$.

\begin{figure}[H]
    \centering
    \vspace{0mm}
    \includegraphics[width=0.5\columnwidth]{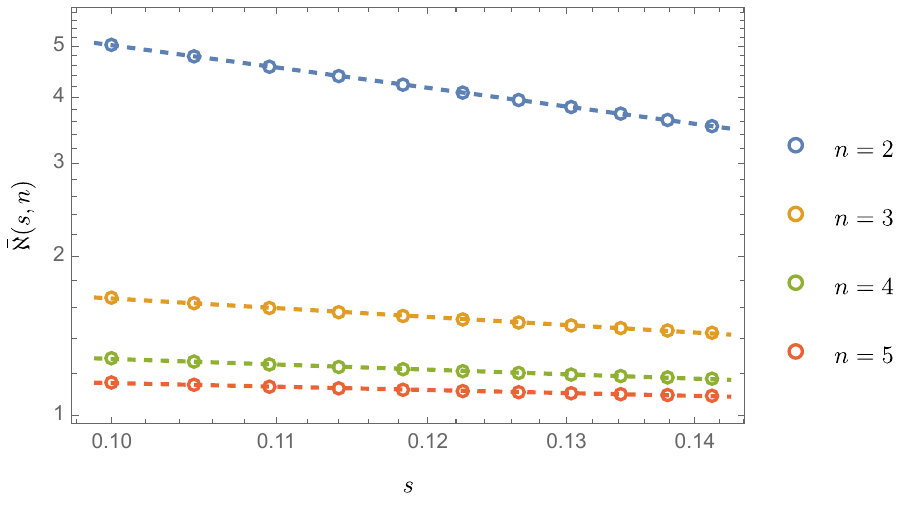}
    \caption{Log-log plots of $\bar{\aleph}(s,n)$ as function of $s$ for different values of $n$. Circles correspond to numerical computations. Dashed lines are fitted curves for the ansatz $\bar{\aleph}(s,n)= \frac{a(n)}{s^{b(n)}}$.}\label{fig:aleph_bar_loglog}
\end{figure}

To test this proposal and find the form of the functions $a(n), b(n)$, we compute $\bar{\aleph}(s,n)$ for small values of $s$ such that $0.01\leq s^{2}\leq0.02$ and different values of $n$. This is done numerically using the same methodology as described in section \ref{Large q limit of the deformed SYK with general n}. The results for $n=2, 3, 4, 5$ are shown in logarithmic scale in figure \ref{fig:aleph_bar_loglog}. The linear form of the plots supports the ansatz in equation (\ref{small s}).
 
\begin{figure}[H]
        \centering
         \subfigure[]{
                \includegraphics[scale=0.5]{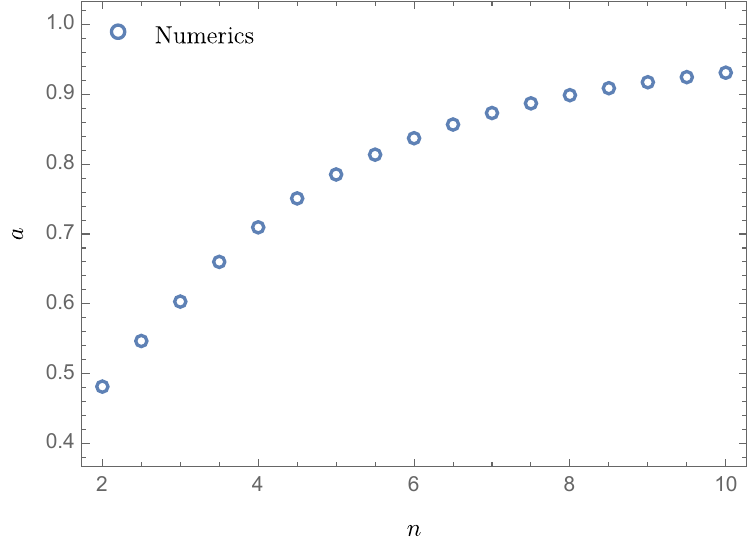}\label{fig:small_s_a}}  \quad\quad
        \subfigure[]{
                \includegraphics[scale=0.5]{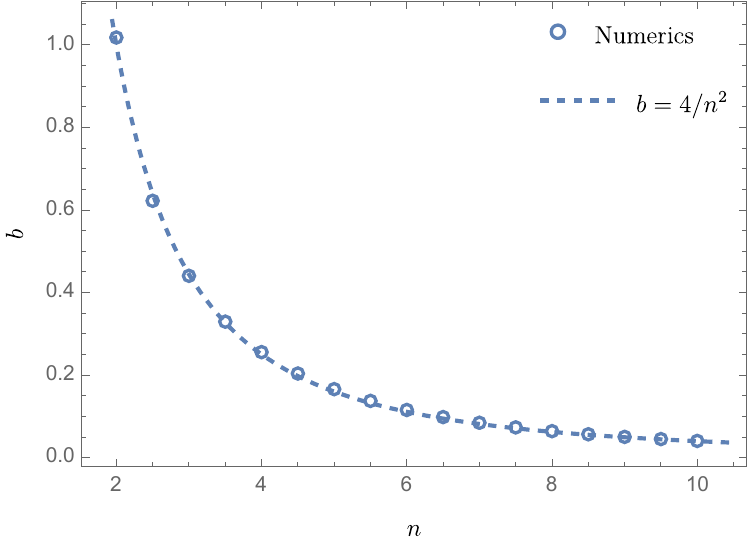} \label{fig:small_s_b}} 
                 \caption{(a) Fitted values for $a(n)$ in the ansatz $\bar{\aleph}(s,n)=\frac{a(n)}{s^{b(n)}}$ for small $s$. (b) Fitted values for $b(n)$ in the ansatz $\bar{\aleph}(s,n)=\frac{a(n)}{s^{b(n)}}$ for small $s$.}
\label{fig:a_b_fit}
\end{figure}

For each $n$, we perform a fit on the data to find $a(n)$ and $b(n)$. For $n=2$, we find $a(n=2) = 0.482$ and $b(n = 2) = 1.02$, which are close to the analytic values of $1/2$ and $1$, respectively. We repeat the procedure for $2 \leq n \leq 10$. The results for $a(n)$ and $b(n)$ are shown in figure \ref{fig:a_b_fit}.

Note that as $n \to \infty$, $a(n \to \infty) \to 1$ and $b(n \to \infty) \to 0$, so $\bar{\aleph} (s,n \to \infty) \to 1$, as expected from the considerations in section \ref{Large q limit of the deformed SYK with general n}. Moreover, a simple fit in figure \ref{fig:small_s_b}, shows that $b(n) \approx 4/n^2$. We conclude that 
\begin{equation}
\label{aleph for small s empirical}
    \bar{\aleph}(s,n) \to \frac{a(n)}{s^{4/n^2}} + \cdots \quad\textrm{as}\quad s\to0 ~,
\end{equation}
with $1/2 \leq a(n) \leq 1$. 

\section{Details on the numerical algorithm to solve the SD equations}
\label{appendix: numerical}
In this appendix we outline the numerical procedure used to solve the Schwinger-Dyson equations (\ref{deformed SD equations}) and (\ref{deformed SD equations ii}). The procedure is analogous to the one described in Appendix G of \cite{Maldacena:2016hyu} for the single SYK model. The idea is to start with the free solution of the single SYK as an initial seed for an iterative algorithm that has fast convergence properties. 

For the numerical procedure, it is convenient to write (\ref{deformed SD equations}) in frequency space, so that at finite temperature the Schwinger-Dyson equations can be written as
\begin{eqnarray}\label{numerical SD equations}
   \frac{1}{G(\omega_n)} &=& -i\omega_n - \Sigma(\omega_n)~,\\
   \Sigma(\tau) &=&\mathcal{J}^2\left(\frac{2^{q-1}}{q} G(\tau)^{q-1} + s^{2} \frac{2^{\tilde{q}-1}}{\tilde{q}} G(\tau)^{\tilde{q}-1}\right) ~. \label{numerical SD equations ii}
\end{eqnarray}
where $\omega_n = 2\pi \left(n+1/2\right)/\beta$ are Matsubara frequencies and $\beta$ is the inverse temperature. At each step in the procedure we update $G(\omega_n)$ by a proportion of the error in \eqref{numerical SD equations},
\begin{equation}
\label{weighted update}
G_{j+1}(\omega_n) = G_{j}(\omega_n) + a \left(\frac{1}{-i\omega_n - \Sigma_{j}(\omega_n)} - G_j(\omega_n)\right)~,
\end{equation}
where the weight $a$ is initially set to $a=0.5$. We then use \eqref{numerical SD equations ii} to get an update for $\Sigma(\omega_n)$, using the fast Fourier transform (FFT) to switch between frequency and position space. The iteration is continued until the error in \eqref{numerical SD equations} is deemed to be sufficiently small. We implemented the algorithm in \verb|python|  using the inbuilt \verb|FFT|, and \verb|IFFT| functions from the \verb|NumPy| module.

To get good convergence, it is important to discretise the $\tau$-interval into many points, particularly near $0$ and $\beta$ where we found the most error from the expected solution. 20,000 points is enough to see good plots, but we could go up to 2,000,000 and still run the algorithm in reasonable time. This allowed us to reach inverse temperatures of the order of $\beta \mathcal{J} \sim 10^{2}$. To reach much larger $\beta \mathcal{J}$ requires significant time and memory.  

Another important aspect in the numerical code is to keep track of the full absolute error squared, $\sum_{n} |G_{j+1}(\omega_n)- G_{j}(\omega_n)|^{2}$,  at each iteration. In the case it increases, we half the value of the weighting parameter $a$. We found that around 50 iterations was sufficient to get convergence to the solution.

\section{Schwarzian action and entropy for the $q=2$ SYK model}
\label{app:Schwarzian contribution to the specific heat}

In this appendix we use the methodology to derive the Schwarzian action employed in sections \ref{Schwarzian for the deep IR} and \ref{Schwarzian for the intermediate IR}  to correctly reproduce the linear-in-temperature entropy of the $q=2$ SYK model at large $N$, which is known to be integrable.

For $q=2$, we can solve the Schwinger-Dyson equations \eqref{field equations large N} and \eqref{field equations large N ii} exactly to find that at low temperatures \cite{Maldacena:2016hyu}
\begin{equation}\label{q=2 entropy}
    \left. \frac{S}{N} \right|_{q=2}= \frac{\pi}{6}\frac{1}{\beta\mathcal{J}} + \cdots~.
\end{equation}
Note that the zero-temperature entropy vanishes for $q=2$. We want to derive this formula from a Schwarzian action perspective. For that, we take $\Sigma \to \Sigma + \partial_{\tau}$  in \eqref{G Sigma action} and write $I = I_{\text{CFT}} + I_{\text{UV}}$ \cite{Kitaev:2017awl, Rosenhaus:2018dtp}, where
\begin{eqnarray} \label{app:CFT action single SYK}
    I_{\text{CFT}}&=&-\frac{1}{2}\log\det(-\Sigma) + \frac{1}{2}\int_{0}^{\beta}\int_{0}^{\beta} d\tau_1 d\tau_2 \left(\Sigma G - \mathcal{J}^2\frac{2^{q-1}}{q^2}G^{q} \right)~,\\
    I_{\text{UV}}&=&\frac{1}{2}\int_{0}^{\beta}\int_{0}^{\beta} d\tau_1 d\tau_2\delta(\tau_{1}-\tau_2)\partial_{\tau_2} G ~.\label{app:pert action intermediate IR}
\end{eqnarray}
We then make an expansion of the saddle solution to $I_{\text{CFT}}$ in powers of $\tau_{12}$. It can be written in terms of soft modes $f(\tau_+)$, see (\ref{Scwharzian expansion intermediate}). We can substitute this expansion into $I_{UV}$, which now becomes an integral over $\tau_{12}$ and $\tau_+$.
Carrying out the $\tau_{12}$ integral with a short time scale cutoff $\varepsilon/\mathcal{J}$, we are left with the following Schwarzian action,
\begin{equation}
    I_{\text{Sch}} = \left[\left(\frac{b (1-q)\varepsilon ^{1-2\Delta}}{6 q^2}\right)\frac{1}{\mathcal{J}} \right]\int_{0}^{\beta}\;d\tau_{+} \;\textrm{Sch}(f(\tau_+),\tau_+) = \left[\left(-\frac{1}{24\pi}\right)\frac{1}{\mathcal{J}}\right] \int_{0}^{\beta}\;d\tau_{+} \;\textrm{Sch}(f(\tau_+),\tau_+) ~,
\end{equation}
where for the last equality we used that $b=1/\pi$ and $\Delta = 1/2$, in the $q=2$ model. Note that the cut off dependence drops out. 
Upon taking this Schwarzian action on-shell, we obtain that the entropy becomes
\begin{equation}\label{q_2 sch entropy}
   \left.  \frac{S_{\text{Sch}}}{N} \right|_{q=2} = \frac{\pi}{6}\frac{1}{\beta\mathcal{J}}~,
\end{equation}
which correctly reproduces \eqref{q=2 entropy}.

\bibliographystyle{JHEP}
\bibliography{bibliography}

\end{document}